\documentclass{arxiv-icarus}

\begin{document}

\begin{frontmatter}
\title{Photometrically-corrected global infrared mosaics of Enceladus:\\New implications for its spectral diversity and geological activity}

\author[LPG]{Rozenn Robidel}\contact{rozenn.robidel@gmail.com}
\author[LPG]{St\'{e}phane Le Mou\'{e}lic}
\author[LPG]{Gabriel Tobie}
\author[LPG]{Marion Mass\'{e}}
\author[JPL]{\\Beno\^{i}t Seignovert}
\author[JPL]{Christophe Sotin}
\author[IPG]{S\'{e}bastien Rodriguez}

\address[LPG]{Laboratoire de Plan\'{e}tologie et G\'{e}odynamique, UMR 6112, CNRS, Universit\'{e} de Nantes, Nantes, France}
\address[JPL]{Jet Propulsion Laboratory, California Institute of Technology, Pasadena CA, USA}
\address[IPG]{Universit\'{e} de Paris, Institut de Physique du Globe de Paris, CNRS, Paris, France}

\begin{abstract}
Between 2004 and 2017, spectral observations have been gathered by the Visual and Infrared Mapping Spectrometer (VIMS) on-board Cassini \citep{Brown2004} during 23 Enceladus close encounters, in addition to more distant surveys. The objective of the present study is to produce a global hyperspectral mosaic of the complete VIMS data set of Enceladus in order to highlight spectral variations among the different geological units. This requires the selection of the best observations in terms of spatial resolution and illumination conditions. We have carried out a detailed investigation of the photometric behavior at several key wavelengths (\num{1.35},\num{1.5}, \num{1.65}, \num{1.8}, \num{2.0}, \num{2.25}, \num{2.55} and \SI{3.6}{\um}), characteristics of the infrared spectra of water ice. We propose a new photometric function, based on the model of \cite{Shkuratov2011} When combined, corrected mosaics at different wavelengths reveal heterogeneous areas, in particular in the terrains surrounding the Tiger Stripes on the South Pole and in the northern hemisphere around \ang{30}N, \ang{90}W. Those areas appear mainly correlated to tectonized units, indicating an endogenous origin, potentially driven by seafloor hotspots.
\end{abstract}

\begin{keyword}
Enceladus \sep Cassini \sep VIMS \sep Spectrophotometry \sep Image processing
\DOI{10.1016/j.icarus.2020.113848}
\end{keyword}

\end{frontmatter}


\section{Introduction}

During its 13 years orbital tour around Saturn, the Cassini spacecraft has acquired a significant amount of observations of Saturn's icy satellites, including Enceladus, an active icy moon only \SI{500}{km} in diameter. Enceladus has an unusually high reflectance due to a surface composed mostly of pure water ice \citep{Buratti1984b, Cruikshank2005, Brown2006, Dalton2010}. Early observations from the Voyager era \citep{Smith1982, Squyres1983} suggest that Enceladus experienced recent resurfacing and volcanic activity. In 2005, after a first hint from the Cassini magnetometer \citep{Dougherty2006}, various instruments of the Cassini spacecraft identified the presence of active jets of water vapor and ice grains emanating from four main warm faults at the South Pole, named Tiger Stripes \citep{Hansen2006, Porco2006, Spencer2006, Waite2006}. This discovery confirms that Enceladus is volcanically active and is the main source of Saturn's E ring, the farthest ring from the planet \citep{Spahn2006}.

The Visual and Infrared Mapping Spectrometer (VIMS) onboard the Cassini spacecraft \citep{Brown2004} provided compositional evidence of activity at Enceladus' South Pole. The analysis of VIMS data provides not only information about the composition of the surface but also the physical state (grain size and degree of crystallinity) of the surface materials and the temperature of the surface \citep{Jaumann2008, Taffin2012, Filacchione2016}. Crystalline water ice with grain size \SI{~100}{\um} as well as \ce{CO2} and light organics were found along the Tiger Stripes \citep{Brown2006, Jaumann2008, Newman2008, Scipioni2017}. \cite{Combe2019} provided the first complete map of \ce{CO2} of Enceladus' surface. They confirmed that \ce{CO2} is mostly located in the south polar terrain, as \ce{CO2} ice and in complexed form, possibly \ce{CO2} clathrate hydrate \citep{Oancea2012}. They also noticed possible detection of \ce{CO2} in the northern hemisphere around \ang{60}N, \ang{270}W and around the North Pole between \ang{70}N and \ang{80}N (absorption band depth at \SI{4.24}{\um}) as well as around 30--\ang{40}N, \ang{100}W (absorption band depth at \SI{2.7}{\um}). These findings need further investigation as only one of the two absorption bands characteristics of \ce{CO2} was observed. \cite{Scipioni2017} investigated the variation of the main water ice absorption bands and of the sub-micron ice grains spectral features, enhancing the Tiger Stripes. Furthermore, they pointed out an abnormally bright behavior around \ang{30}N, \ang{90}W which does not seem, a priori, to be related to any obvious morphological structure. Interestingly, \cite{Ries2015} detected a microwave scattering anomaly covering the same area.\\

In \cite{Combe2019} and \citep{Scipioni2017}, no photometric correction was applied during the processing of the spectra. In the present study, we investigate this aspect, to refine the spectral analysis and compositional mapping. Indeed, the brightness of an object depends on the intrinsic properties of the surface materials (composition, grain size, roughness, porosity etc.) and the surface shape but also on the geometry of illumination and observation. To construct surface reflectance maps from VIMS hyperspectral cubes taken at varying geometries, photometric analyses are to be achieved. It is also needed for comparing surface spectral reflectance from one region to another observed under different geometries and interpreting the composition based on laboratory measurements, taken at geometries different from the planetary observations. To correct the surface reflectance from the geometry effects, several photometric models have been developed \citep[\eg][]{Hapke1963, Hapke1981, Hapke2012, Minnaert1941, Shkuratov2011}. Here we adapt the approach of \cite{Shkuratov2011} which has been applied on other Saturnian moons by \cite{Filacchione2018a, Filacchione2018}.
In this work, we carry a detailed photometric analysis in order to derive an optimal photometric function of Enceladus' surface at different wavelengths. In sections \ref{sec:sec_2} and \ref{sec:sec_3}, we present the parametrization of the photometric function and the construction of the global mosaics. This parametrization takes into account various viewing and illumination conditions which allows us to produce global mosaics of the complete dataset of Enceladus acquired by Cassini VIMS during the entire mission. We then discuss in \sref{sec_4} the spectral diversity and the implications of these new compositional mappings for the geological activity of Enceladus.

\section{VIMS dataset and methods for computing global maps}\label{sec:sec_2}
\subsection{Observations and data selection}\label{sec:sec_2.1}

VIMS acquired hyperspectral cubes composed of images taken at 352 spectral channels ranging from \num{0.35} to \SI{5.12}{\um} \citep{Brown2004}. The instrument consisted of two imaging spectrometers. The first one operated in the visible range (\SIrange{0.35}{1.04}{\um}) over \num{96} spectral channels whereas the second one covered a wavelength range from \num{0.88} to \SI{5.12}{\um} over \num{256} spectral channels. Images were up to \num{64 x 64} pixels, the size depending on the operation mode (image, line, point or occultation) and observation conditions (fast flyby, particular geometry, simultaneous observation with the Imaging Science Subsystem - ISS).

All the data cubes have been calibrated in radiance factor \citep{Hapke1981}:

\vspace{-.3cm}
\begin{equation}
    r_F = \pi r
\end{equation}

Where the $r$ is the bidirectional reflectance of the surface. It is defined as follows:

\vspace{-.15cm}
\begin{equation}
    r(i, e, \alpha, \lambda) = \frac{I(i, e, \alpha, \lambda)}{J(\lambda)}
\end{equation}

\vfill\null

Where $I$ is the radiance in \si{\watt\per\square\meter\per\um\per\steradian}, $J$ is the normal solar irradiance in \si{\watt\per\square\meter\per\um}, $\lambda$ is the wavelength, $i$ is the local angle of incidence, $e$ is the local angle of emergence and $\alpha$ is the phase angle. The radiance factor is also known as $I/F$ (with $F = J / \pi$). We use the term \emph{reflectance} for the radiance factor in the remainder of this paper.

To convert digital count number of each pixel into physical radiance, the calibration pipeline (background subtraction, flat fielding, conversion into specific energy, division by the solar spectrum and despiking) used is RC19 \citep{Clark2018}. Due to the spectral drift in wavelength of the spectral channel, data are realigned on a common spectral base with a spline interpolation, using the shifts evaluated by \cite{Clark2018}. Navigation information, such as geographic position (latitude and longitude position at the pixel's center and at the four corners of the pixel), corresponding illumination and viewing conditions (incidence, emergence and solar phase angles at each pixel's center) and spatial resolution, were retrieved using reconstructed SPICE kernels \citep{Acton1996}. The pixel location is calculated as the intersection between Enceladus mean radius (\SI{252.1}{km}) and the VIMS pixel boresight \citep{Nicholson2019}. Our projected VIMS cubes are geometrically consistent with the latest ISS global map \citep{Bland2018} as it can been on a few examples on \figref{fig_11}.

During the 13 years of the Cassini mission, 147 flybys of Enceladus were performed, 23 of which were targeted. We reported a total of \num{14742} hyperspectral cubes acquired throughout the mission, \num{2486} of which were out of scope or pointing at the night side, thus not relevant for the surface mapping.

\begin{figure*}[!ht]
    \includegraphics[width=.9\linewidth]{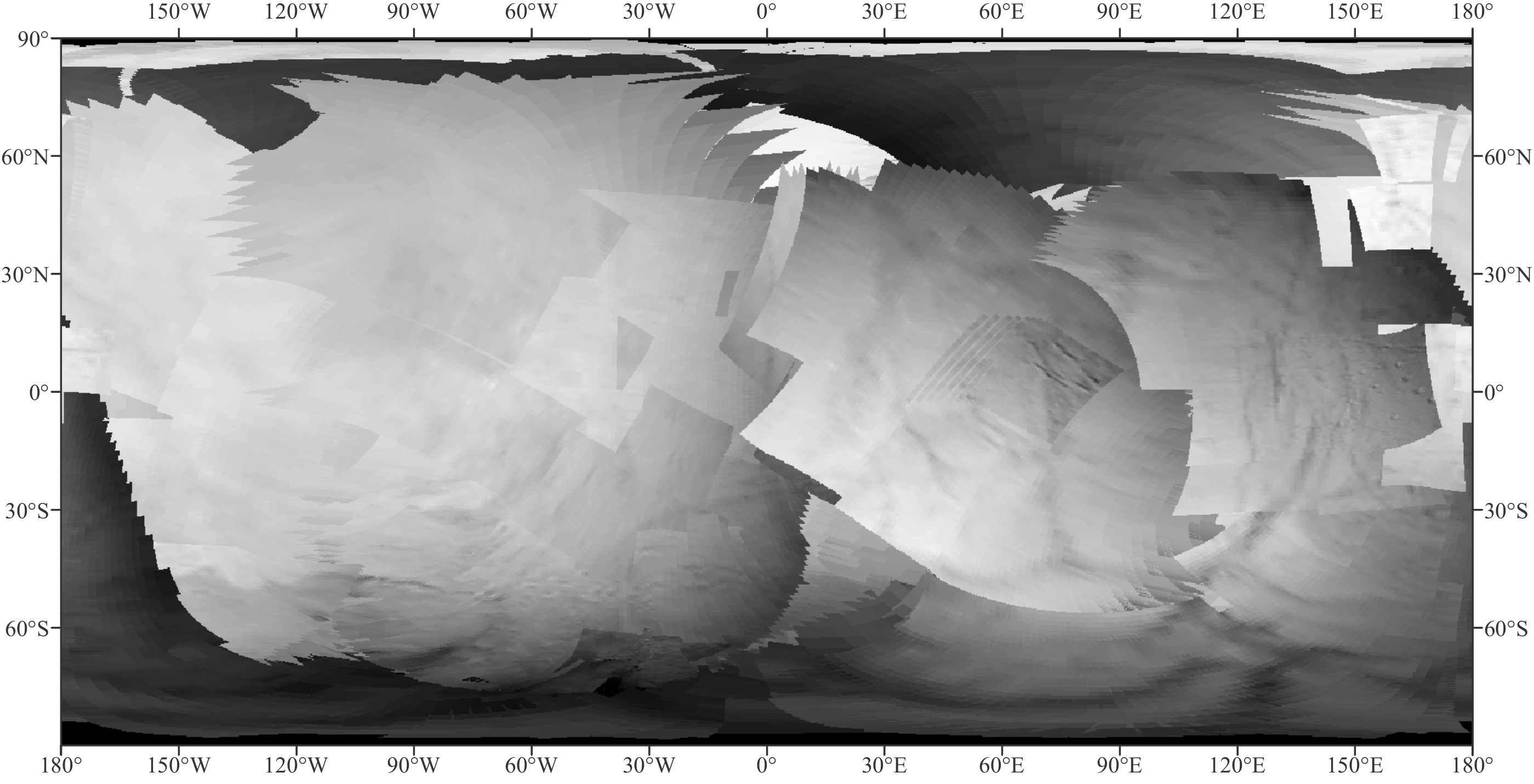}
    \caption{Global mosaic in equirectangular projection at \SI{1.8}{\um}, containing all the 355 VIMS cubes responding to thresholds mentioned in \sref{sec_2.1}, before the photometric correction.}
    \label{fig:fig_1}
\end{figure*}

\begin{figure*}[!ht]
    \vspace{.25cm}
    \includegraphics[width=.9\linewidth]{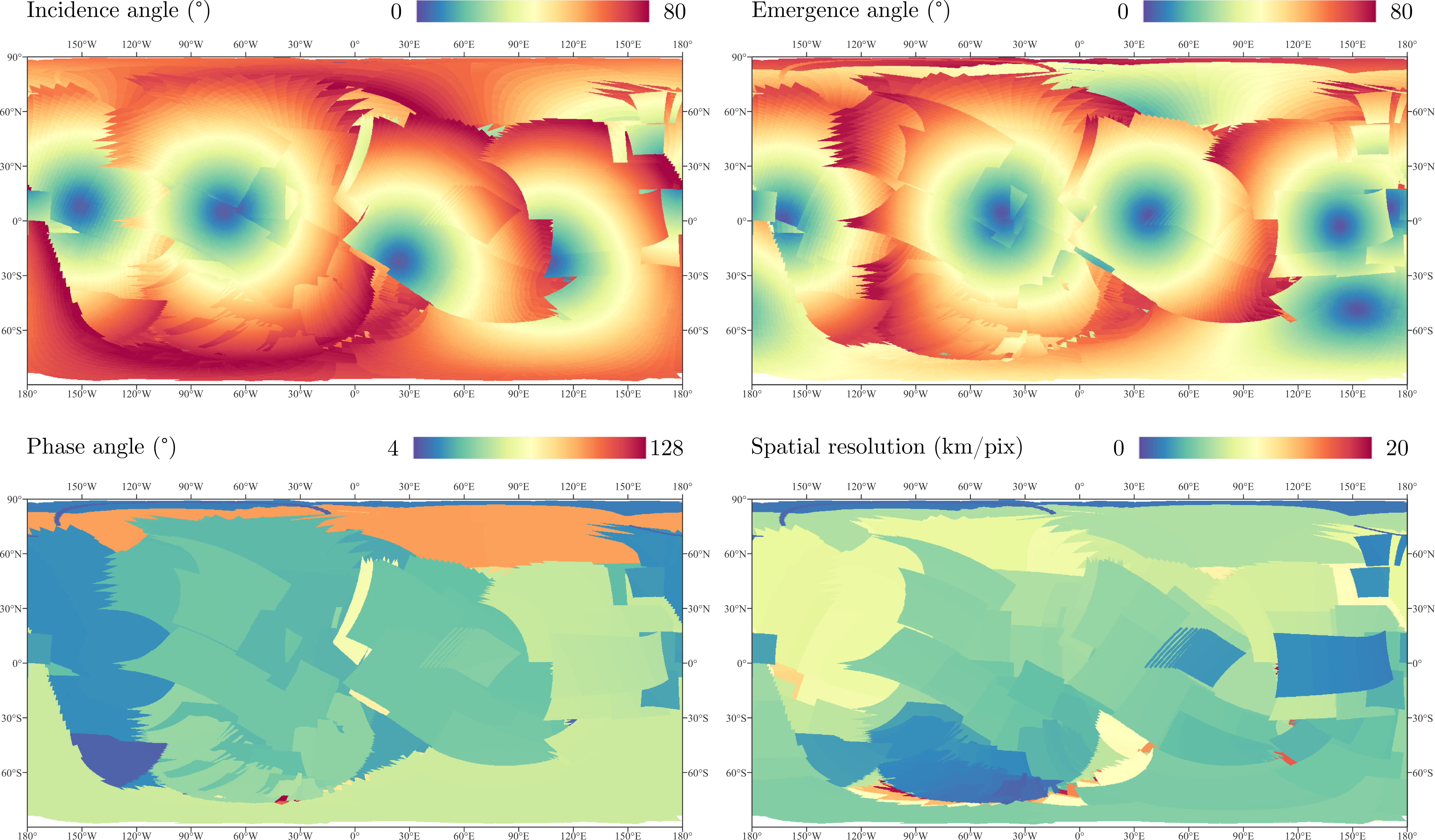}
    \caption{Global mosaics of incidence, emergence, phase angles and the pixel spatial resolution.}
    \label{fig:fig_2}
\end{figure*}

The imaging data acquired on planetary surfaces strongly depends on the angles at which they are observed. These conditions vary according to the orbit, the seasons and the sun's illumination. When the acquisition geometry is too extreme, \eg very high angles or grazing observation conditions, data are difficult to reconcile with each other into homogeneous mosaics. Unfavorable observation geometry is generally not suitable to produce global maps. To avoid those unfavorable observation geometries, we have applied thresholds on the incidence and emergence angles. We considered only observations acquired on the dayside of Enceladus. We have performed several tests and reached a compromise between spatial coverage and mosaic quality by limiting the local angles of incidence and emergence to \ang{80}. In addition, we consider only hyperspectral cubes with a spatial resolution better than \SI{20}{km/pixel} in order to have the most resolved dataset. Finally, cubes which are partially or totally saturated are not considered. To discard unusable cubes, we use the vims website \href{https://vims.univ-nantes.fr}{vims.univ-nantes.fr} which provides a user-friendly access to multispectral summary images, produced from Planetary Data System archive and covering the entire mission.

When restrained to these conditions, the number of useful VIMS cubes is reduced to 355, acquired during the most favorable flybys. They come from revolutions 3, 4, 11, 88, 91, 120, 121, 131, 136, 141, 142, 153, 158, 223, 228, 230 and 250.

\subsection{Merging data into global maps}

Global maps are generated by sorting the 355 cubes in decreasing order of spatial resolution, which puts the most resolved images on top of the map. This method has been successfully applied to Titan \citep{LeMouelic2019}. The resulting global mosaic, including 355 selected cubes (\figref{fig_1}), shows high contrasted seams and illumination residuals due to the lack of an appropriate photometric correction.

Although we did not apply any threshold on the solar phase angle, the imposed limitations on the incidence and emergence angles have resulted in phase angle ranging from \ang{4} to \ang{128} (\figref{fig_2}). As shown in \figref{fig_2}, illustrating the relevant observation parameters, the observation conditions can locally strongly vary from one flyby to another. This effect can be partially corrected by applying an appropriate photometric function. We will focus on the derivation of this function in the next step.

\section{Derivation of a parametrized photometric correction function}\label{sec:sec_3}
\subsection{Model description}

The application of a photometric correction is an essential step in obtaining a map expressing the surface variability in terms of composition and physical state. In fact, the observed brightness of the surface depends on both illumination and viewing geometry.

To model the photometric behavior, different models have been proposed \citep{Minnaert1941, Buratti1985a, Shkuratov2011}, one of the most commonly used being the model of \cite{Hapke1963, Hapke1981, Hapke2012}. Previous studies of Enceladus' photometry only dealt with Voyager and Earth-based data, hence with low resolution \citep{Buratti1984, Buratti1984b, Verbiscer1994}. \cite{Buratti1984}, tested an empirical expression and Minnaert's model \citep[eqs. (1) and (2) in][]{Buratti1984}, while \cite{Verbiscer1994} tested Hapke's model.

In this study, we have decided to adopt the method given by \cite{Shkuratov2011}. The photometric correction is therefore separated in two parts; the disk function $D(i, e, \alpha)$ and the phase function $A_\textrm{eq} (\alpha, \lambda)$ that includes the albedo:

\begin{equation}
    I/F (i, e, \alpha, \lambda) = D(i, e, \alpha) \cdot A_\textrm{eq} (\alpha, \lambda)
\end{equation}

Similar approach has been previously applied on the Earth's Moon \citep{Shkuratov1999, Shkuratov2011, Kreslavsky2000}, Mercury \citep{Domingue2016}, Vesta \citep{Schroder2013, Combe2015}, Ceres \citep{Schroder2017}, Tethys and Dione \citep{Filacchione2018a, Filacchione2018}.

We have evaluated different disk functions \citep{Minnaert1941, Akimov1976, Akimov1988}. We also have tested the case of parametrized Akimov \citep[eq. (19) in][]{Shkuratov2011}. The models' equations and results are provided in supplementary material (\figref{fig_S1}, \figref{fig_S2} and \tabref{tab_S1}). The standard deviation errors of Akimov model and parametrized Akimov model, associated with a linear phase, are fairly close. However, the Akimov model provides the best result in terms of mosaic quality, therefore, in the remainder of this paper, we use Akimov disk function, defined as follows:

\begin{equation}\small\label{eq:eq_4}
    D(i, e, \alpha) = \cos\left( \frac{\alpha}{2} \right)
    \cdot
    \cos\left[
        \frac{\pi}{\pi - \alpha} \left( \gamma - \frac{\alpha}{2} \right)
    \right]
    \cdot
    \frac{\left( \cos\beta \right)^{ \frac{\alpha}{\pi - \alpha} }}{\cos\gamma}
\end{equation}

The photometric latitude $\beta$ and longitude $\gamma$ are related to the local angles of incidence and emergence as follows:

\begin{eqnarray}\label{eq:eq_5}
    \cos i &=& \cos\beta \cdot \cos(\alpha - \gamma) \\
    \cos e &=& \cos\beta \cdot \cos\gamma
\end{eqnarray}

The phase function, or equigonal albedo, $A_\textrm{eq} (\alpha, \lambda)$, represents the phase dependence of brightness. It can be expressed as follows:

\begin{equation}\label{eq:eq_6}
    A_\textrm{eq} (\alpha, \lambda) = A \left( \alpha_0, \lambda \right) \cdot f (\alpha, \lambda)
\end{equation}

Where $f (\alpha, \lambda)$ is the phase function normalized to unity at $\alpha = \alpha_0 = \ang{0}$ and $A \left( \alpha_0, \lambda \right)$ is the equigonal albedo value at $\alpha = \alpha_0 = \ang{0}$ \citep{Shkuratov2011}. The phase function is derived by applying a fit of arbitrary functional form to the observation data corrected for the disk function. We have tested both exponential fit and polynomial fits of various degrees and noticed that a linear fit is satisfactory. Hence, the following linear fit is used:

\begin{equation}\label{eq:eq_8}
    A_\textrm{eq} (\alpha, \lambda) = a (\lambda) + b (\lambda) \cdot \alpha
\end{equation}

We can notice that the parameter a corresponds to the equigonal albedo value at $\alpha = \ang{0}$.

A few examples showing that the equigonal albedo varies linearly with the phase angle are given in the next section (\figref{fig_4}).

\subsection{Phase curve fitting at selected wavelengths}\label{sec:sec_3.2}

In the range \SIrange{0.88}{5.12}{\um}, water ice exhibits strong features (\num{1.04}, \num{1.25}, \num{1.5}, \num{1.65}, \num{2.0} and \SI{3.1}{\um}). The existence, position, shape, intensity and width of absorption bands are characteristics of the structure of the ice as well as its temperature and grain size \citep{Fink1975, Clark1984, Brown2006, Newman2008, Clark2013, Scipioni2017}. The deeper the absorption, the higher the abundance and the grain size. The reflectance peak at \SI{3.6}{\um} is also an indicator of the grain size, the greater the intensity of the peak, the smaller the grain size will be \citep{Hansen2004, Filacchione2012}. The \SI{1.65}{\um} absorption band and the \SI{3.1}{\um} Fresnel reflection peak are the two most obvious indicators of crystallinity, both of which are much prominent in crystalline ice \citep{Schmitt1998, Brown2006, Newman2008}. Note that the \SI{1.65}{\um} absorption band is also sensitive to the water ice temperature, being deeper for colder temperatures of crystalline ice \citep{Grundy1998, Grundy1999}.

\begin{figure}[!ht]
    \includegraphics[width=.9\linewidth]{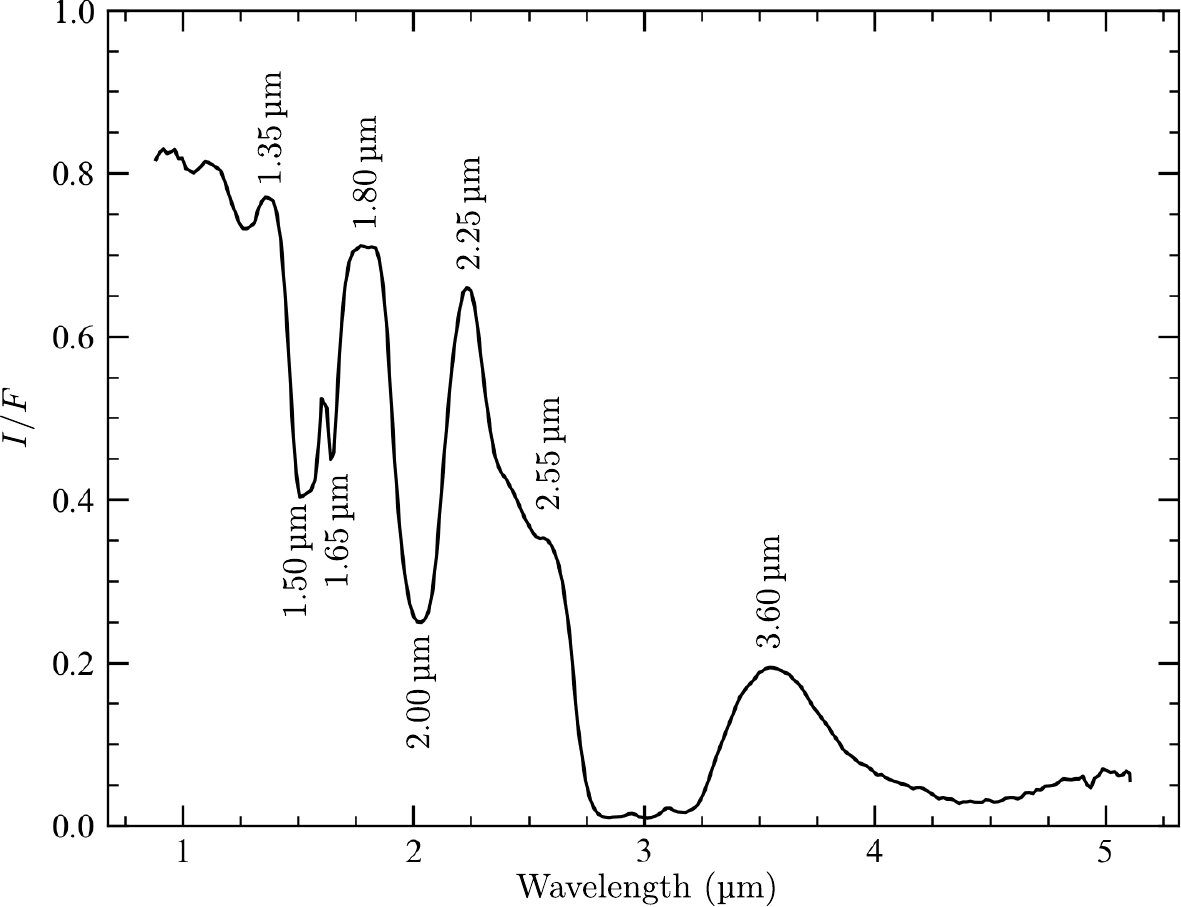}
    \caption{Typical $I/F$ spectrum of Enceladus surface, exhibiting the characteristic absorption bands of water ice. Parameters of observation: $i = \ang{39}$, $e = \ang{50}$ and $\alpha = \ang{15}$.}
    \label{fig:fig_3}
\end{figure}

In our study, we have focused our efforts on eight wavelengths (\figref{fig_3}), including absorption bands at \num{1.5}, \num{1.65} and \SI{2.0}{\um} and the reflectance peak at \SI{3.6}{\um}. We have chosen not to use the Fresnel reflection peak to derive our photometric function, as data are often too noisy around this wavelength and are thus not suitable for our study. The other selected wavelengths (\ie \num{1.35}, \num{1.8}, \num{2.25} and \SI{2.55}{\um}) are intentionally located outside characteristic water ice absorption bands. At wavelengths longer than \SI{3}{\um}, the reflectance is in general noisier than at shorter wavelengths due to intrinsic low reflectance of the surface and a lower instrumental sensitivity. Therefore, we did not use wavelengths larger than \SI{3.6}{\um}.

In order to keep a good signal to noise ratio, we filtered Enceladus' data by selecting pixels having $I/F \le 0.01$ for each selected wavelength. We have also applied thresholds for each pixel on the values of local angles of incidence and emergence and spatial resolution (see \sref{sec_2.1}). Pixels occurring at the limb and intersecting only part of the surface have been removed.

\begin{figure}[!ht]
    \includegraphics[width=.9\linewidth]{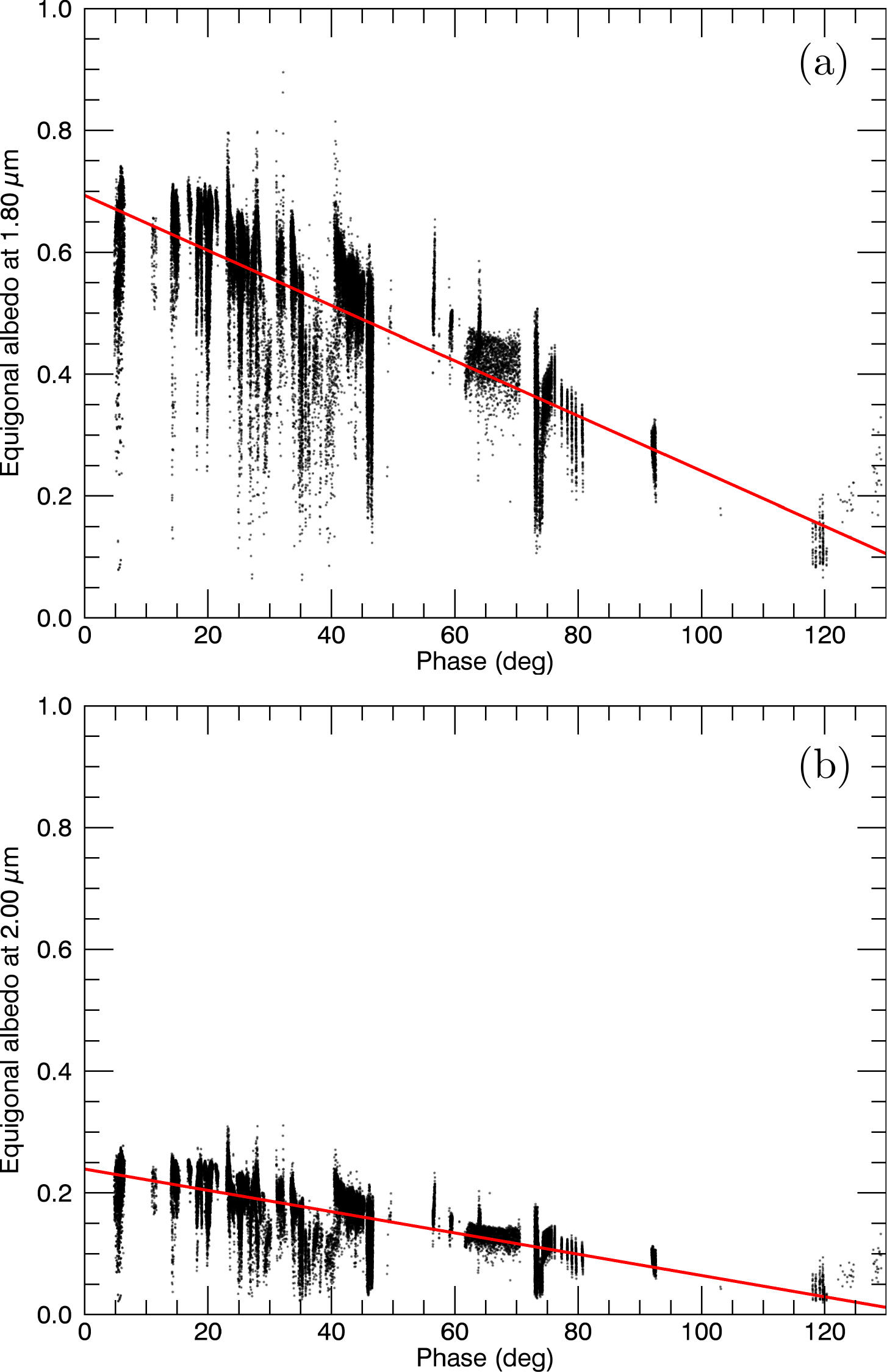}
    \caption{Enceladus' phase curves at \SI{1.8}{\um} (a) and \SI{2.0}{\um} (b). The red curves correspond to the fit computed on the entire dataset for $(i, e) < \ang{80}$ and $\alpha < \ang{130}$.}
    \label{fig:fig_4}
\end{figure}

As a result of this filtering, we first correct data for the disk function (\eqref{eq_4}). We then plot the dependence of the equigonal albedo with the phase (\figref{fig_4}).

We notice variability along the vertical axis. It partly reflects a diverse variety of terrains, associated with different shadowing effects. The significant scatter around \ang{30} of phase corresponds to cubes with a kilometric resolution, which therefore contains a contribution of spatially resolved shadows. This variability is also partly related to pixels occurring near the limb and/or terminator. We did not filter these observations as, despite their extreme observing geometries, they allow an improvement of the spatial coverage and are among the data with the best spatial resolution.

We then apply a linear fit (\eqref{eq_8}) to the data corrected for the disk function (\eqref{eq_4}), thus obtaining parameters a and b, listed in \tabref{tab_1}. The diversity of the values illustrates the spectral dependence of the surface behavior.

\begin{table}[!ht]
    \caption{Photometric fit parameters, valid for $(i, e) < \ang{80}$ and $\alpha < \ang{130}$.}
    \label{tab:tab_1}
    \begin{tabular}{c c c}
    \toprule
    Wavelength (\si{\um}) & $a$ & $b$ (\si{\per\radian}) \\
    \midrule
    1.3595 & 0.771 & -0.268 \\
    1.5079 & 0.394 & -0.156 \\
    1.6567 & 0.483 & -0.193 \\
    1.8040 & 0.698 & -0.250 \\
    2.0017 & 0.242 & -0.098 \\
    2.2495 & 0.638 & -0.226 \\
    2.5644 & 0.333 & -0.121 \\
    3.5961 & 0.186 & -0.085 \\
    \bottomrule
    \end{tabular}
\end{table}

\subsubsection{Generalizing the phase function}

\begin{figure}[!ht]
    \includegraphics[width=.9\linewidth]{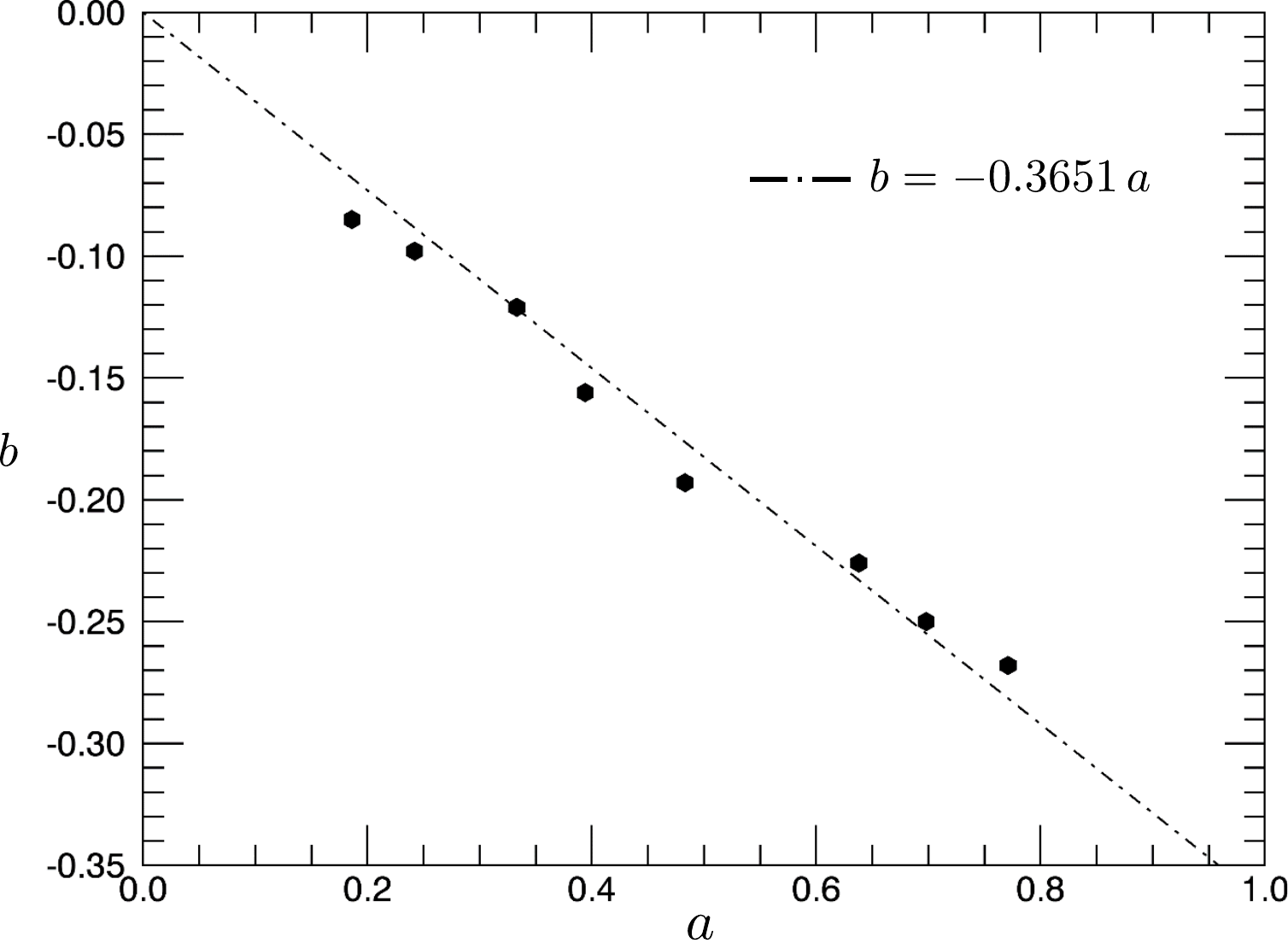}
    \caption{Correlation between phase function parameters $a$ and $b$.}
    \label{fig:fig_5}
\end{figure}

Our objective is to retrieve parameters for any wavelength, including those often affected by a low signal to noise ratio, such as around the Fresnel peak (\SI{3.1}{\um}), which is characteristic of the crystallinity of the ice (see \sref{sec_3.2}). Once the parameters a and b are computed at the eight selected key wavelengths, we notice that they strongly correlate: the ratio $b / a$ is nearly constant (\figref{fig_5}).

The final photometric model we have derived in this work is the following:

\begin{equation}\scriptsize\label{eq:eq_9}
    I/F
    =
    \cos\left( \frac{\alpha}{2} \right)
    \cdot
    \cos\left[
        \frac{\pi}{\pi - \alpha} \left( \gamma - \frac{\alpha}{2} \right)
    \right]
    \cdot
    \frac{\left( \cos\beta \right)^{ \frac{\alpha}{\pi - \alpha} }}{\cos\gamma}
    \cdot
    a \cdot \left( 1 - 0.37 \alpha \right)
\end{equation}

\vfill\null

Where $\alpha$ is the phase angle, $\gamma$ the photometric longitude, $\beta$ the photometric latitude and a the zero phase angle equigonal albedo. The photometric latitude and longitude are related to the local angles of incidence and emergence (see \eqref{eq_5} and \eqref{eq_6}). The angles are expressed in radians. It should be noted that the model only depends on the illumination and viewing conditions and the wavelength.

\subsection{Application of the new photometric function to global mosaics}

\begin{figure}[!ht]
    \includegraphics[width=.9\linewidth]{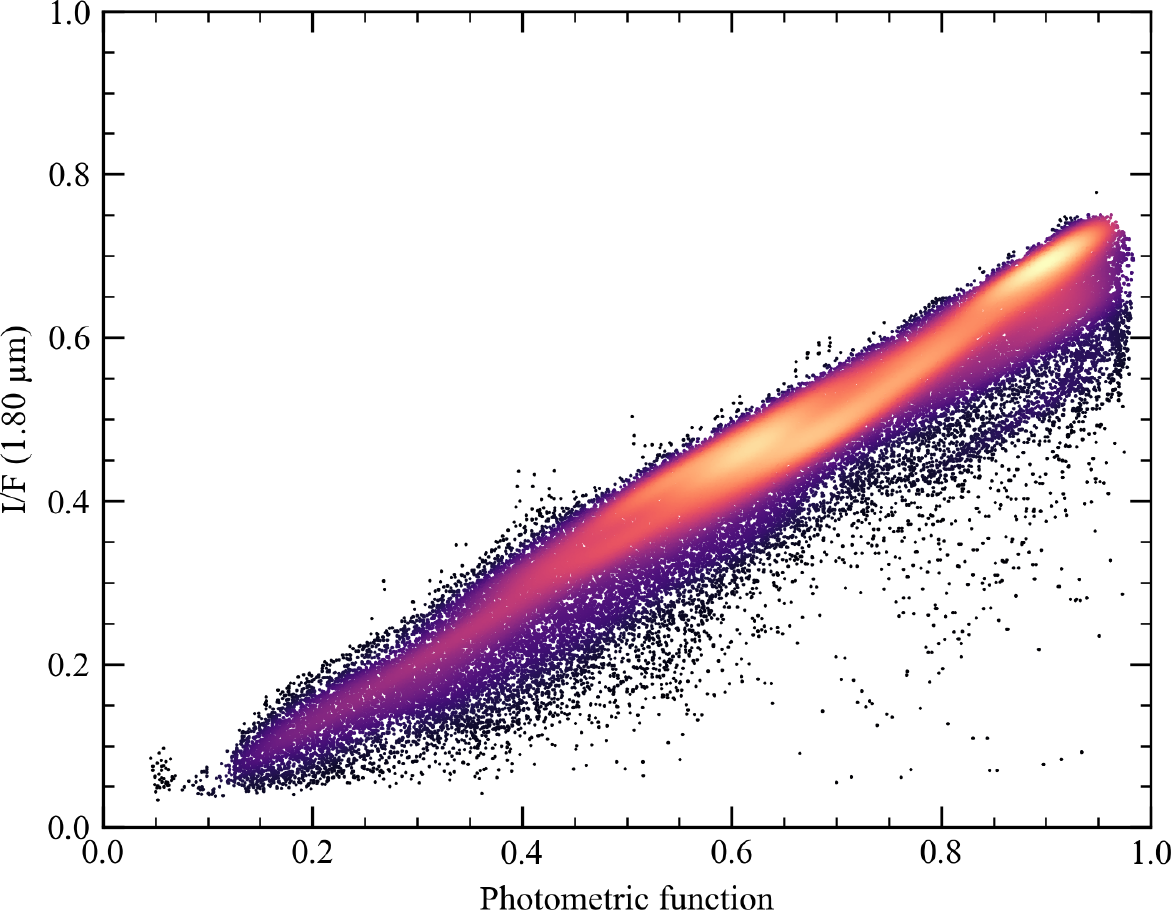}
    \caption{I/F at \SI{1.8}{\um} versus the derived photometric function (see \eqref{eq_9}).}
    \label{fig:fig_6}
\end{figure}

In this section, we focus on one of our selected wavelengths (\SI{1.8}{\um}), located outside absorption bands characteristic of water ice. \figref{fig_6} shows the correlation between the reflectance at \SI{1.8}{\um} and the photometric function we have derived (\eqref{eq_9}). The linear behavior confirms that the photometric function corrects at first order for the incidence, emergence and phase variations. Data out of the main trend, representing less than \SI{3}{\percent} of the total data, correspond mostly to pixels with resolved shadows, or located along the Tiger Stripes or occurring near the limb and/or terminator.

\begin{figure*}[!ht]
    \includegraphics[width=.9\linewidth]{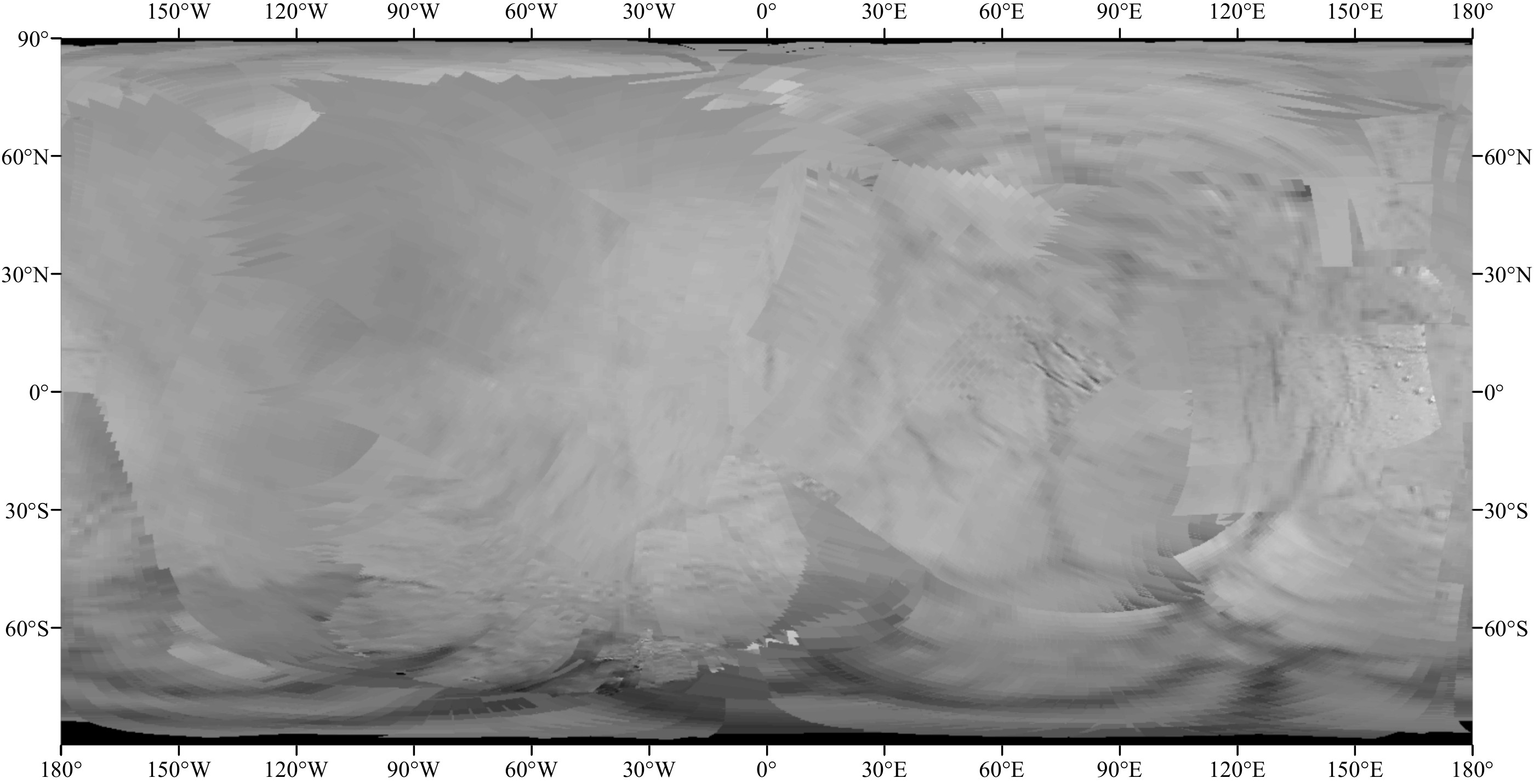}
    \caption{Equigonal albedo map at \SI{1.8}{\um} normalized to unity at zero phase angle.}
    \label{fig:fig_7}
\end{figure*}

The global map at \SI{1.8}{\um} corrected for the photometry with the function described in \eqref{eq_9} is shown in \figref{fig_7}. This photometric correction allows a significant reduction of the level of seams in almost all regions. The remaining ones seem to be related to pixels occurring having extreme observing conditions. Other global maps, derived from various other photometric tests, are provided in the supplementary material (\figref{fig_S1} and \figref{fig_S2}). The association of the Akimov disk function with a linear phase function was the best trade-off we were able to obtain.

\begin{figure*}[!ht]
    \vspace{.25cm}
    \includegraphics[width=.84\linewidth]{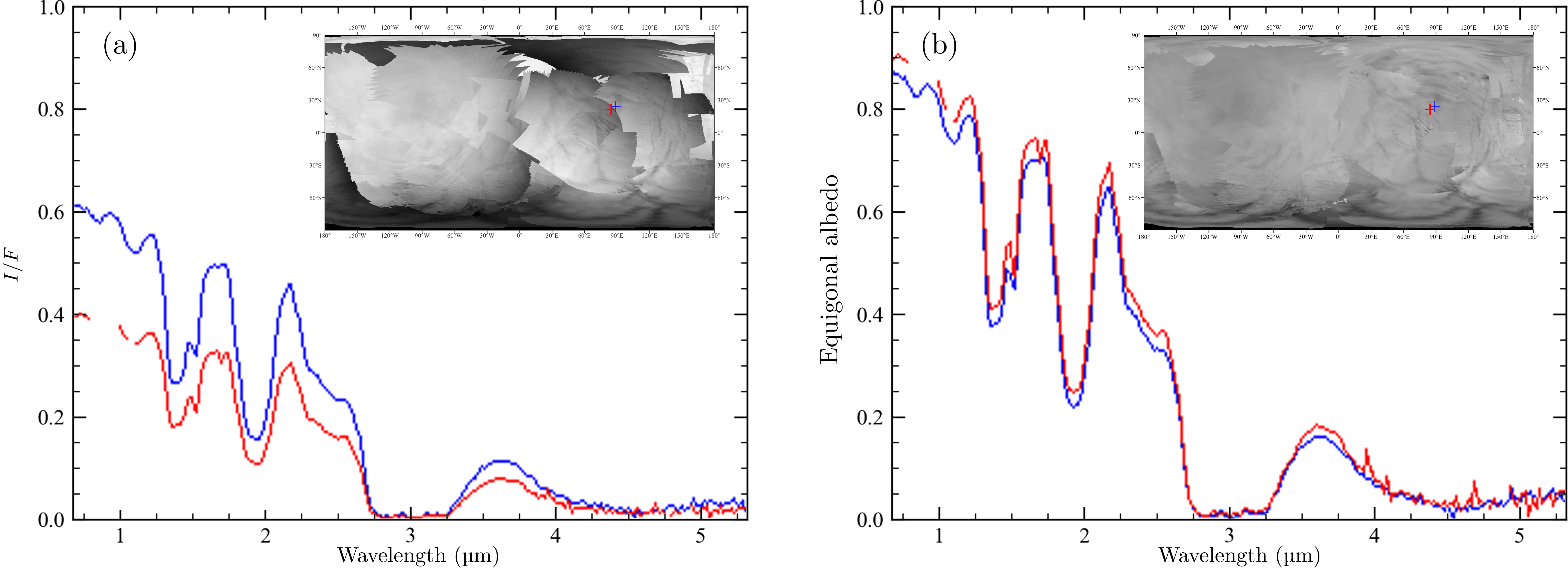}
    \caption{Spectrum comparison before (a) and after (b) the photometric correction. The red spectrum corresponds to the cube 1487299582\_1 located at \ang{22.5}N, \ang{85}E ($i = \ang{72}$, $e = \ang{49}$, $\alpha = \ang{26}$) while the blue spectrum corresponds to the cube 1489049741\_1 located at \ang{23.5}N, \ang{88}E ($i = \ang{50}$, $e = \ang{65}$, $\alpha = \ang{45}$).}
    \label{fig:fig_8}
\end{figure*}

As we have computed the phase function parameters at 8 wavelengths and used an interpolation to extrapolate the correction to the 256 spectral channels, we can obtain corrected reflectance spectra. To better illustrate the effect of the photometric correction, we compare spectra of two points close to each other located in a region with a significant seam on the mosaic (blue and red crosses on \figref{fig_8}a) before photometric correction. The first is located at \ang{22.5}N, \ang{85}E while the second is located at \ang{23.5}N, \ang{88}E. The geometry of observation and illumination is very different from a point to another. \figref{fig_8} compares these spectra before (\figref{fig_8}a) and after (\figref{fig_8}b) the photometric correction. Spectra are similar after the photometric correction, revealing a homogenous terrain which is confirmed by the ISS map. The photometric correction thus allows to compare spectra independently of illumination and observation conditions.

\begin{figure*}[!ht]
    \includegraphics[width=.78\linewidth]{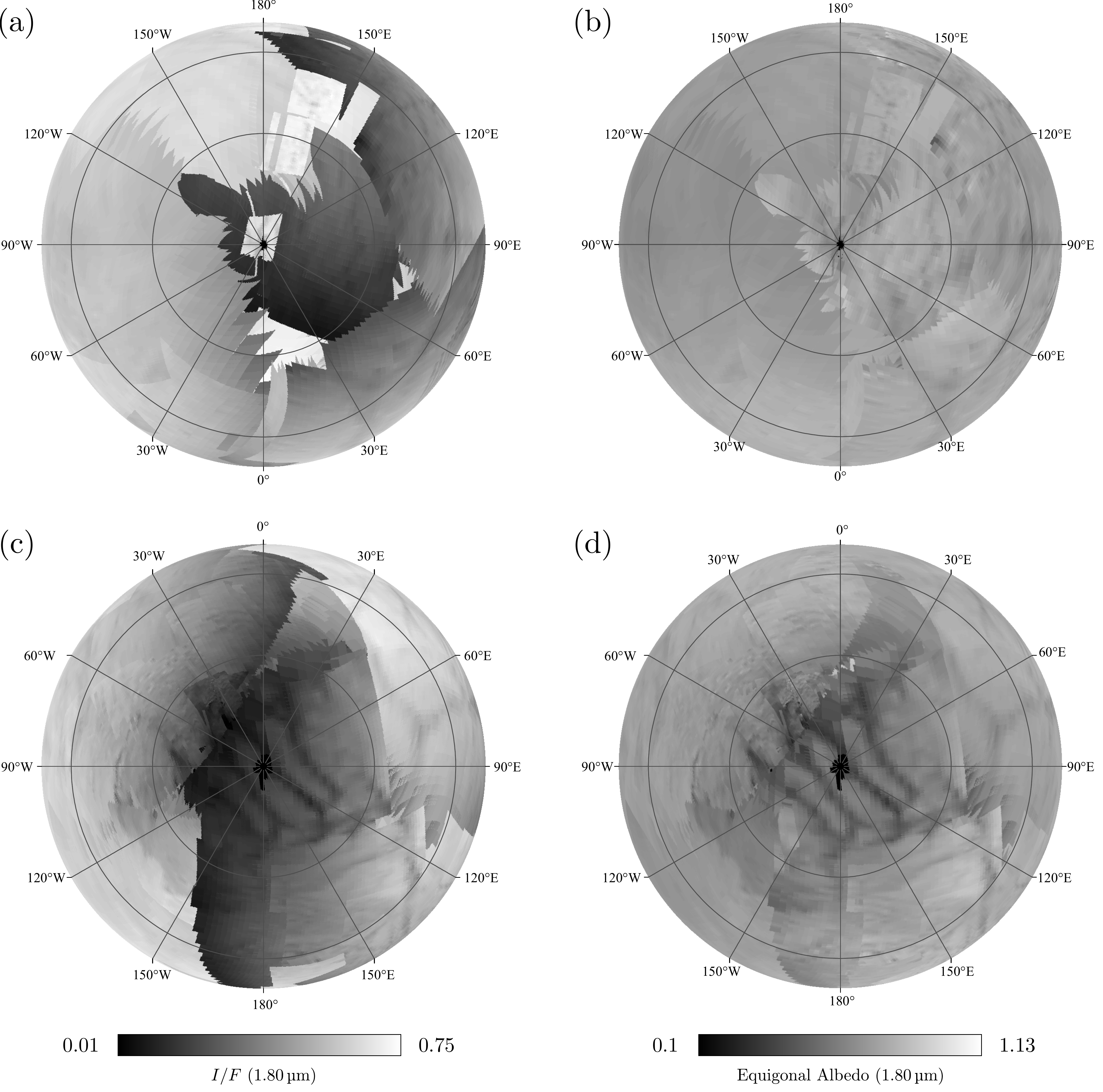}
    \caption{Uncorrected mosaic at \SI{1.8}{\um} in orthographic projection centered on the North Pole (a) and the South Pole (c). \SI{1.8}{\um} map corrected with the photometric function described in \eqref{eq_9} centered on the North Pole (b) and the South Pole (d).}
    \label{fig:fig_9}
\end{figure*}

\figref{fig_9} shows a comparison of the map before and after the photometric correction in orthographic projection centered on the North Pole and the South Pole. We can see that the application of the new photometric correction reveals features and particularly emphasizes brightness variations across the Tiger Stripes.

\section{Spectral diversity}\label{sec:sec_4}

To investigate the spectral diversity of Enceladus' surface, we have computed RGB false color composites of photometrically corrected mosaics, assuming a specific set of color channel attributions. The red, green and blue channels correspond respectively to the ratio between 3.1 and \SI{1.65}{\um} bands, \SI{2.0}{\um} and \SI{1.8}{\um}. These wavelengths have been selected to emphasize variations of the ice physical state. We deliberately use a mix of a ratio (red) and single bands (green and blue) to keep the dependence with the albedo patterns while still emphasizing color variations. As aforementioned, both \num{3.1} and \SI{1.65}{\um} bands are indicators of the ice crystallinity (see \sref{sec_3.2}). The ratio between the two bands thus enhances the variations due to ice crystallinity, as \SI{3.1}{\um} is the Fresnel reflection peak while the \SI{1.65}{\um} is an absorption band. The \SI{2.0}{\um} channel represents the center of a strong water ice absorption band, while the \SI{1.8}{\um} channel falls in the continuum outside of water ice bands. For the composites, we have computed a median of 3 spectral channels (\ie 134, 135 and 136 at \SI{3.08}{\um}, \SI{3.10}{\um} and \SI{3.11}{\um}) for the Fresnel peak to be less dependent on noise.

\begin{figure*}[!ht]
    \vspace{.2cm}
    \includegraphics[width=.78\linewidth]{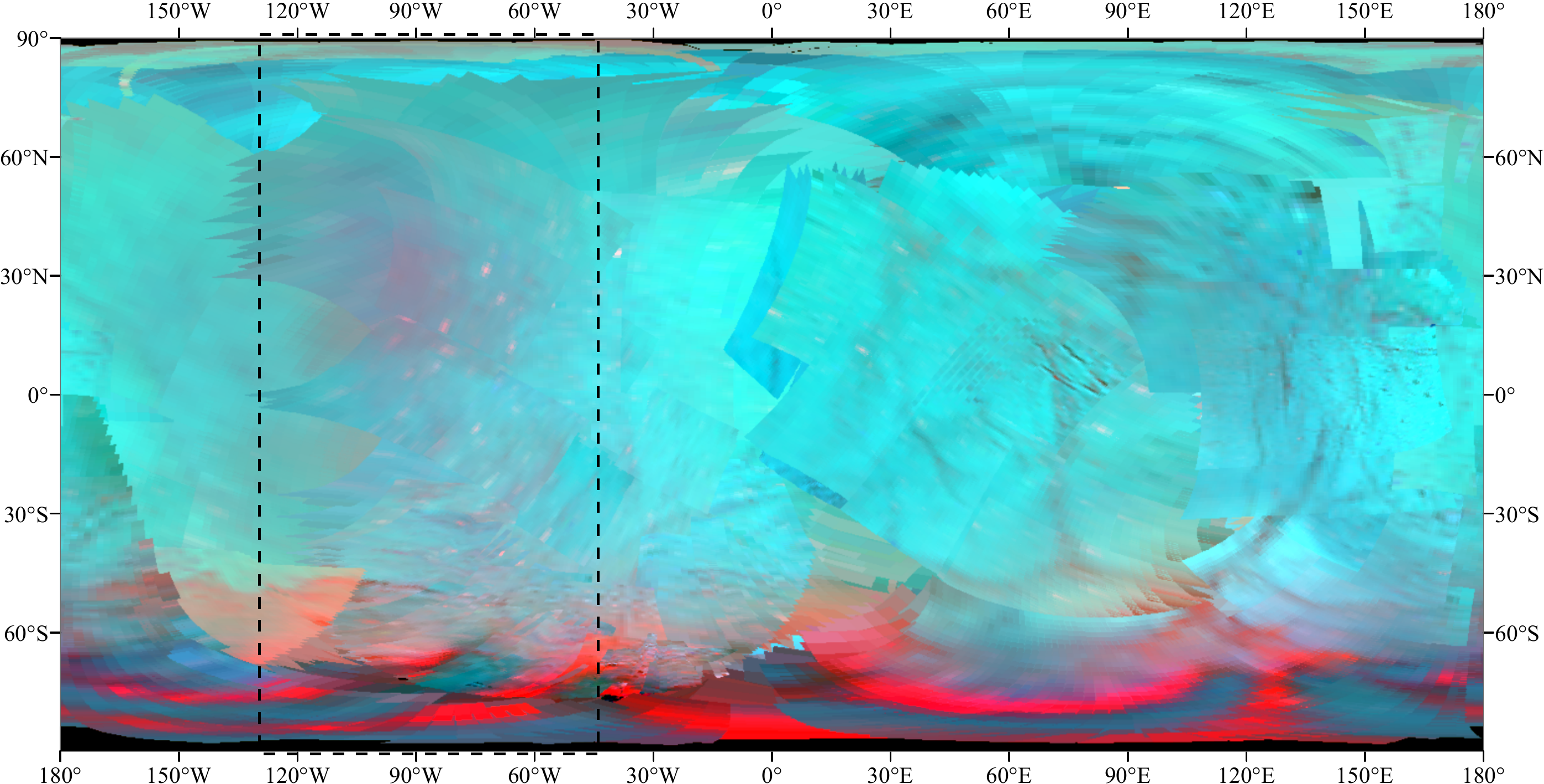}
    \caption{RGB global map in equirectangular projection with the red, green and blue channels controlled by the \SI{3.1/1.65}{\um} ratio, \SI{2.0}{\um} and \SI{1.8}{\um} channels respectively. The black dashed line rectangle corresponds to an area discussed in \figref{fig_12}.}
    \label{fig:fig_10}
\end{figure*}

\begin{figure*}[!ht]
    \includegraphics[width=.65\linewidth]{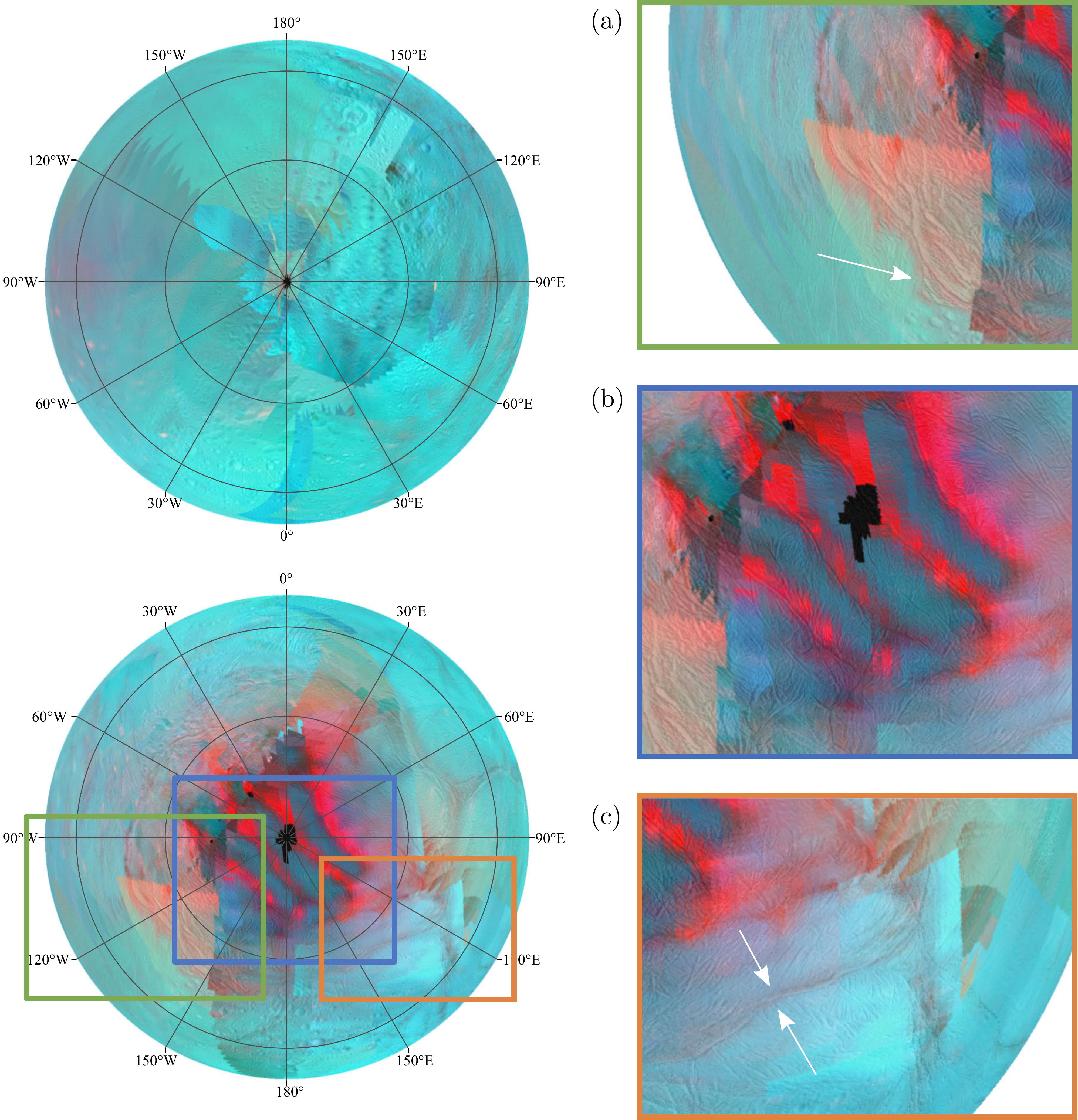}
    \caption{Orthographic views derived from the RGB corrected color map (\figref{fig_10}) overlapped on ISS mosaic produced by \cite{Bland2018}. The top left image is centered at Enceladus' North Pole whereas the bottom left one is centered at Enceladus' South Pole. Images (a), (b) and (c) show zooms of Enceladus' South Pole.}
    \label{fig:fig_11}
\end{figure*}

Consistent with the results of \cite{Brown2006}, we observe a strong enhancement of the red color along the Tiger Stripes (\figref{fig_11}b), revealing the highest degree of crystallinity (highest \SI{3.1}{\um} peak and most pronounced \SI{1.65}{\um} absorption band, resulting in the highest \num{3.1/1.65} ratio, see \figref{fig_10} and \figref{fig_11}). Interestingly, this RGB color composite reveals a clear boundary in the circumpolar terrain between the leading hemisphere and southern curvilinear terrains as defined by \cite{Crow-Willard2015} (white arrows on \figref{fig_11}a and \figref{fig_11}c). This suggests that spectral variations in these areas are primarily controlled by geological processes and not mainly driven by particle deposits as proposed by previous studies \citep{Schenk2011, Scipioni2017}.

\begin{figure}[!ht]
    \includegraphics[width=.9\linewidth]{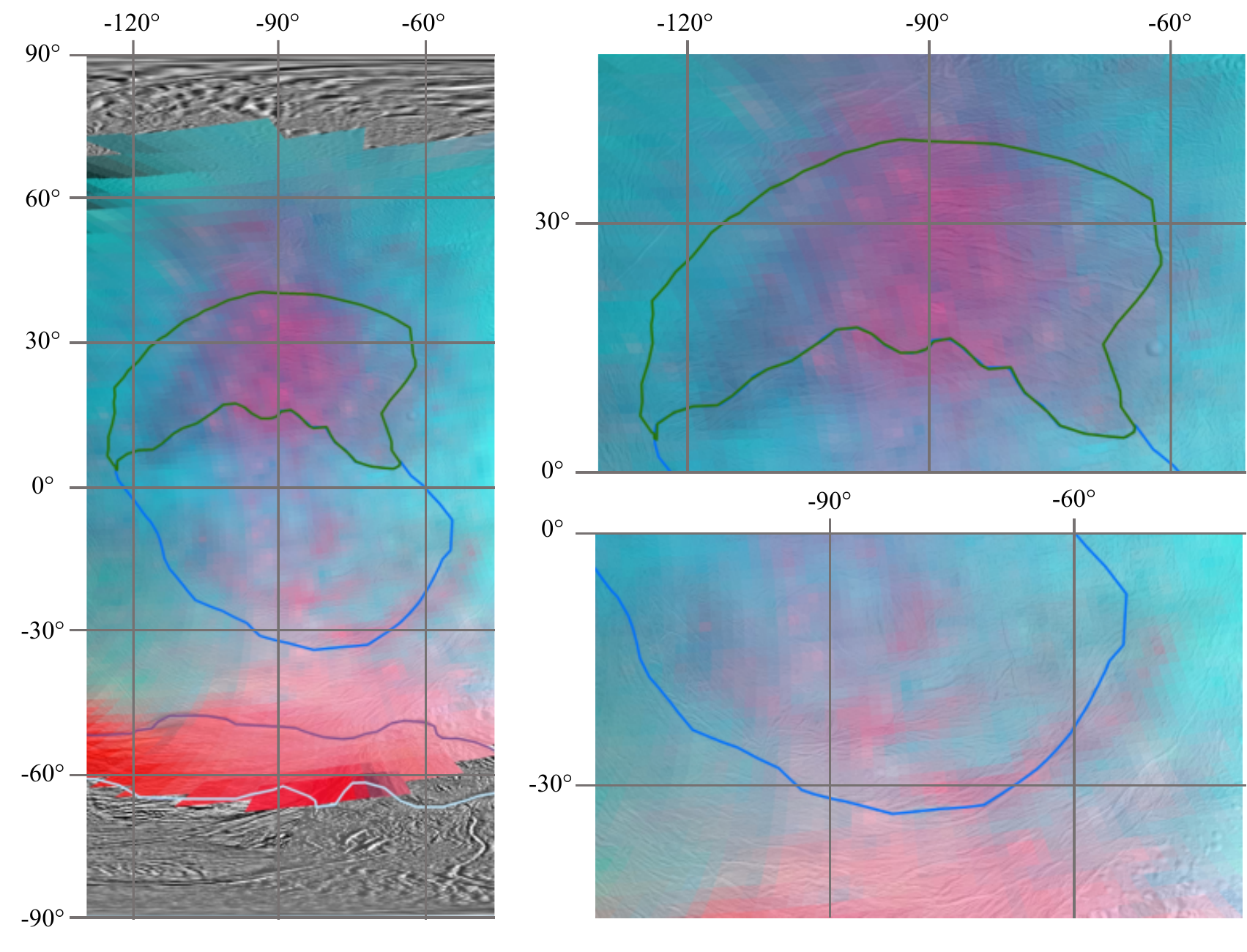}
    \caption{Correlation between our VIMS RGB composite representation, corresponding to the black rectangle on \figref{fig_10} and the geological units as defined by \cite{Crow-Willard2015}. The green line delimits the Leading Hemisphere smooth unit, the blue one delimits the central Leading Hemisphere terrain, the purple one delimits the southern curvilinear unit whereas the cyan one delimits the central south polar unit. The RGB composite was applied on one specific cube (cube: 1702359174\_1). We have superimposed the result on ISS mosaic by \cite{Bland2018}.}
    \label{fig:fig_12}
\end{figure}

Note that residual bright red dots cannot be unambiguously related to any particular spectral feature. They likely correspond to noise, all the more so as they do not really show up in the cube displayed on \figref{fig_12}, which is acquired with a long-time exposure (therefore with a better signal to noise ratio). As already mentioned in \sref{sec_3.2}, the wavelengths longer than \SI{3}{\um} are generally more affected by the noise.

The RGB composite shown in \figref{fig_10} contains a red area centered around \ang{30}N, \ang{90}W which was previously identified by different studies \citep{Ries2015, Scipioni2017, Combe2019}. As the global map in \figref{fig_10} still shows some residual seams, we decided to use a single cube in this specific map of \figref{fig_12} in order to minimize the photometric effects and to also check the consistency of the spectral signature. We show that this area is well correlated with the Leading Hemisphere smooth unit defined by \cite{Crow-Willard2015}. Notably, the southern boundary of this unit coincides relatively well with the limit of the reddish area. We also notice a few reddish zones between \ang{15} and \ang{30}S which appear correlated with major trough features.

The corrected spectral map indicates that the main spectral variations are well correlated with several tectonized terrains, suggesting that these variations are mostly controlled by endogenous processes. The reddish zones are indicative of enhanced ice crystallinity, implying that fresh water ices have been exposed geologically recently. It is difficult to really constrain the age of the different reddish terrains, but we can at least conclude that, outside the south polar terrain, the Leading Hemisphere smooth unit centered at \ang{30}N and \ang{90}W is the youngest terrain. Its longitudinal location is consistent with the predictions of \cite{Choblet2017} concerning preferential seafloor hotspot locations along the \ang{90} meridian. Compared to seafloor hotspots located in polar regions, \cite{Choblet2017} predicted that hotspots occurring at mid-latitude along this meridian are more chaotic in nature, and do not remain stable more than a few million years. The reddish spectral behavior of this area indicates that it has likely been active in a recent past, potentially during a period of higher orbital eccentricity and enhanced tidal dissipation in the porous rock core of Enceladus.

\section{Conclusion}

We have processed the entire VIMS hyperspectral archive of Enceladus in order to compute a global mosaic. The selection of data was based on automatic filters and manual refinement using the previews provided on the website \href{https://vims.univ-nantes.fr}{vims.univ-nantes.fr}. 355 cubes were merged to produce a global map at \SI{16}{pixels} per degree.

The product of this refined dataset required the correction of illumination effects. The disk function is based on the Akimov model. We have derived a new empirical phase function. This new photometric function allows to decrease the level of seams in the mosaics of Enceladus computed from the global VIMS data archive. Some seams are, however, still present. They are mostly due to data acquired in strongly varying observing conditions and thus difficult to reconcile with each other into homogeneous mosaics. Some data having extreme observing conditions (\eg near the limb and/or terminator) were not filtered out as they improve the spatial coverage and provide the best spatial resolution. The study was a true compromise between the quality of the mosaic and the spatial coverage. For future studies, these data could be filtered out in order to improve the derivation of the photometric function. Nonetheless, it will be challenging to completely erase the seams due to strong variability in spatial resolution, pointing conditions and observing geometry. An orbiter around Enceladus would provide a much more optimized cartographic data set compared to the series of distant flybys performed by Cassini.

Furthermore, we produced a new global color map which strongly emphasizes the distribution of the main spectral units, particularly near the Tiger Stripes. The area centered around \ang{30}N, \ang{90}W is correlated with the Leading Hemisphere smooth terrain defined by \cite{Crow-Willard2015}, and may be the signature of recent activity driven by seafloor hotspots, comparable to the south polar activity \citep{Choblet2017}. We plan in further studies to apply this new photometric function to other icy satellites, to evaluate possible differences with Enceladus.

\section*{Acknowledgements}

This work acknowledges the financial support from CNES, based on observations with Cassini/VIMS, as well as in preparation of the ESA JUICE mission, and from R\'{e}gion Pays de la Loire, project GeoPlaNet (convention \#2016-10982). The authors thank R. Pappalardo and A. Patthoff for kindly providing us their structural-geological map of Enceladus. Authors are very grateful to G. Filacchione and S. Schr\"{o}der for their detailed and helpful comments.

\vfill\null


\bibliography{biblio}

\begin{thebibliography}{55}
\providecommand{\natexlab}[1]{#1}
\providecommand{\url}[1]{\texttt{#1}}
\expandafter\ifx\csname urlstyle\endcsname\relax
  \providecommand{\doi}[1]{doi: #1}\else
  \providecommand{\doi}{doi: \begingroup \urlstyle{rm}\Url}\fi

\bibitem[Acton(1996)]{Acton1996}
Acton C.~H.
\newblock \href{http://doi.org/10.1016/0032-0633(95)00107-7}{Ancillary data
  services of {{NASA}}'s navigation and {{Ancillary Information Facility}}}.
\newblock \emph{Planetary and Space Science}, 44\penalty0 65--70, {\bf 1996}.

\bibitem[Akimov(1976)]{Akimov1976}
Akimov L.~A.
\newblock \href{http://adsabs.harvard.edu/abs/1976SvA....19..385A}{Influence of
  mesorelief on the brightness distribution over a planetary disk}.
\newblock \emph{Soviet Astronomy}, 19\penalty0 385--388, {\bf 1976}.

\bibitem[Akimov(1988)]{Akimov1988}
Akimov L.~A.
\newblock \href{http://adsabs.harvard.edu/abs/1988KFNT....4Q...3A}{Light
  reflection by the moon.}
\newblock \emph{Kinematika i Fizika Nebesnykh Tel}, 4\penalty0 3--10, {\bf
  1988}.

\bibitem[Bland et~al.(2018)Bland, Becker, Edmundson, Roatsch, Archinal, Takir,
  Patterson, Collins, Schenk, Pappalardo, and Cook]{Bland2018}
Bland M.~T. and~10~colleagues.
\newblock \href{http://doi.org/10.1029/2018EA000399}{A {{New Enceladus Global
  Control Network}}, {{Image Mosaic}}, and {{Updated Pointing Kernels From
  Cassini}}'s 13-{{Year Mission}}}.
\newblock \emph{Earth and Space Science}, 5\penalty0 (10)\penalty0 604--621,
  {\bf 2018}.

\bibitem[Brown et~al.(2004)Brown, Baines, Bellucci, Bibring, Buratti,
  Capaccioni, Cerroni, Clark, Coradini, Cruikshank, Drossart, Formisano,
  Jaumann, Langevin, Matson, Mccord, Mennella, Miller, Nelson, Nicholson,
  Sicardy, and Sotin]{Brown2004}
Brown R.~H. and~21~colleagues.
\newblock \href{http://doi.org/10.1007/1-4020-3874-7_3}{The {{Cassini Visual}}
  and {{Infrared Mapping Spectrometer}} ({{VIMS}}) {{Investigation}}}.
\newblock In \emph{The {{Cassini}}-{{Huygens Mission}}}, pp 111--168. {Kluwer
  Academic Publishers}, {\bf 2004}.

\bibitem[Brown et~al.(2006)Brown, Clark, Buratti, Cruikshank, Barnes, Mastrapa,
  Bauer, Newman, Momary, Baines, Bellucci, Capaccioni, Cerroni, Combes,
  Coradini, Drossart, Formisano, Jaumann, Langevin, Matson, McCord, Nelson,
  Nicholson, Sicardy, and Sotin]{Brown2006}
Brown R.~H. and~24~colleagues.
\newblock \href{http://doi.org/10.1126/science.1121031}{Composition and
  {{Physical Properties}} of {{Enceladus}}' {{Surface}}}.
\newblock \emph{Science}, 311\penalty0 (5766)\penalty0 1425--1428, {\bf 2006}.

\bibitem[Buratti and Veverka(1984)]{Buratti1984b}
Buratti B. and Veverka J.
\newblock \href{http://doi.org/10.1016/0019-1035(84)90042-3}{Voyager photometry
  of {{Rhea}}, {{Dione}}, {{Tethys}}, {{Enceladus}} and {{Mimas}}}.
\newblock \emph{Icarus}, 58\penalty0 (2)\penalty0 254--264, {\bf 1984}.

\bibitem[Buratti(1984)]{Buratti1984}
Buratti B.~J.
\newblock \href{http://doi.org/10.1016/0019-1035(84)90109-X}{Voyager disk
  resolved photometry of the {{Saturnian}} satellites}.
\newblock \emph{Icarus}, 59\penalty0 (3)\penalty0 392--405, {\bf 1984}.

\bibitem[Buratti and Veverka(1985)]{Buratti1985a}
Buratti B.~J. and Veverka J.
\newblock \href{http://doi.org/10.1016/0019-1035(85)90094-6}{Photometry of
  rough planetary surfaces: {{The}} role of multiple scattering}.
\newblock \emph{Icarus}, 64\penalty0 (2)\penalty0 320--328, {\bf 1985}.

\bibitem[Choblet et~al.(2017)Choblet, Tobie, Sotin, B{\v e}hounkov{\'a}, {\v
  C}adek, Postberg, and Sou{\v c}ek]{Choblet2017}
Choblet G. and~6~colleagues.
\newblock \href{http://doi.org/10.1038/s41550-017-0289-8}{Powering prolonged
  hydrothermal activity inside {{Enceladus}}}.
\newblock \emph{Nature Astronomy}, 1\penalty0 (12)\penalty0 841--847, {\bf
  2017}.

\bibitem[Clark and Lucey(1984)]{Clark1984}
Clark R.~N. and Lucey P.~G.
\newblock \href{http://doi.org/10.1029/JB089iB07p06341}{Spectral properties of
  ice-particulate mixtures and implications for remote sensing: 1. {{Intimate}}
  mixtures}.
\newblock \emph{Journal of Geophysical Research: Solid Earth}, 89\penalty0
  (B7)\penalty0 6341--6348, {\bf 1984}.

\bibitem[Clark et~al.(2013)Clark, Carlson, Grundy, and Noll]{Clark2013}
Clark R.~N. and~3~colleagues.
\newblock \href{http://doi.org/10.1007/978-1-4614-3076-6_1}{Observed {{Ices}}
  in the {{Solar System}}}.
\newblock In \emph{The {{Science}} of {{Solar System Ices}}}, pp 3--46.
  {Springer}, {\bf 2013}.

\bibitem[Clark et~al.(2018)Clark, Brown, Lytle, and Hedman]{Clark2018}
Clark R.~N. and~3~colleagues.
\newblock
  \href{http://atmos.nmsu.edu/data_and_services/atmospheres_data/Cassini/vims.html}{The
  {{VIMS Wavelength}} and {{Radiometric Calibration}} 19, {{Final Report}}}.
\newblock {The Planetary Atmospheres Node}, {\bf 2018}.

\bibitem[Combe et~al.(2015)Combe, Ammannito, Tosi, De~Sanctis, McCord, Raymond,
  and Russell]{Combe2015}
Combe J.-P. and~6~colleagues.
\newblock \href{http://doi.org/10.1016/j.icarus.2015.07.034}{Reflectance
  properties and hydrated material distribution on {{Vesta}}: {{Global}}
  investigation of variations and their relationship using improved calibration
  of {{Dawn VIR}} mapping spectrometer}.
\newblock \emph{Icarus}, 259\penalty0 21--38, {\bf 2015}.

\bibitem[Combe et~al.(2019)Combe, McCord, Matson, Johnson, Davies, Scipioni,
  and Tosi]{Combe2019}
Combe J.-P. and~6~colleagues.
\newblock \href{http://doi.org/10.1016/j.icarus.2018.08.007}{Nature,
  distribution and origin of {{CO2}} on {{Enceladus}}}.
\newblock \emph{Icarus}, 317\penalty0 491--508, {\bf 2019}.

\bibitem[Crow-Willard and Pappalardo(2015)]{Crow-Willard2015}
Crow-Willard E.~N. and Pappalardo R.~T.
\newblock \href{http://doi.org/10.1002/2015JE004818}{Structural mapping of
  {{Enceladus}} and implications for formation of tectonized regions}.
\newblock \emph{Journal of Geophysical Research: Planets}, 120\penalty0
  (5)\penalty0 928--950, {\bf 2015}.

\bibitem[Cruikshank et~al.(2005)Cruikshank, Owen, Ore, Geballe, Roush, {de
  Bergh}, Sandford, Poulet, Benedix, and Emery]{Cruikshank2005}
Cruikshank D.~P. and~9~colleagues.
\newblock \href{http://doi.org/10.1016/j.icarus.2004.09.003}{A spectroscopic
  study of the surfaces of {{Saturn}}'s large satellites: {{H2O}} ice, tholins,
  and minor constituents}.
\newblock \emph{Icarus}, 175\penalty0 (1)\penalty0 268--283, {\bf 2005}.

\bibitem[Dalton et~al.(2010)Dalton, Cruikshank, Stephan, McCord, Coustenis,
  Carlson, and Coradini]{Dalton2010}
Dalton J.~B. and~6~colleagues.
\newblock \href{http://doi.org/10.1007/s11214-010-9665-8}{Chemical
  {{Composition}} of {{Icy Satellite Surfaces}}}.
\newblock \emph{Space Science Reviews}, 153\penalty0 (1)\penalty0 113--154,
  {\bf 2010}.

\bibitem[Domingue et~al.(2016)Domingue, Denevi, Murchie, and
  Hash]{Domingue2016}
Domingue D.~L. and~3~colleagues.
\newblock \href{http://doi.org/10.1016/j.icarus.2015.11.040}{Application of
  multiple photometric models to disk-resolved measurements of {{Mercury}}'s
  surface: {{Insights}} into {{Mercury}}'s regolith characteristics}.
\newblock \emph{Icarus}, 268\penalty0 172--203, {\bf 2016}.

\bibitem[Dougherty et~al.(2006)Dougherty, Khurana, Neubauer, Russell, Saur,
  Leisner, and Burton]{Dougherty2006}
Dougherty M.~K. and~6~colleagues.
\newblock \href{http://doi.org/10.1126/science.1120985}{Identification of a
  {{Dynamic Atmosphere}} at {{Enceladus}} with the {{Cassini Magnetometer}}}.
\newblock \emph{Science}, 311\penalty0 (5766)\penalty0 1406--1409, {\bf 2006}.

\bibitem[Filacchione et~al.(2012)Filacchione, Capaccioni, Ciarniello, Clark,
  Cuzzi, Nicholson, Cruikshank, Hedman, Buratti, Lunine, Soderblom, Tosi,
  Cerroni, Brown, McCord, Jaumann, Stephan, Baines, and
  Flamini]{Filacchione2012}
Filacchione G. and~18~colleagues.
\newblock \href{http://doi.org/10.1016/j.icarus.2012.06.040}{Saturn's icy
  satellites and rings investigated by {{Cassini}}\textendash{{VIMS}}: {{III}}
  \textendash{} {{Radial}} compositional variability}.
\newblock \emph{Icarus}, 220\penalty0 (2)\penalty0 1064--1096, {\bf 2012}.

\bibitem[Filacchione et~al.(2018{\natexlab{a}})Filacchione, Ciarniello,
  D'Aversa, Capaccioni, Cerroni, Buratti, Clark, Stephan, and
  Plainaki]{Filacchione2018a}
Filacchione G. and~8~colleagues.
\newblock \href{http://doi.org/10.1029/2018GL078602}{Photometric {{Modeling}}
  and {{VIS}}-{{IR Albedo Maps}} of {{Tethys From Cassini}}-{{VIMS}}}.
\newblock \emph{Geophysical Research Letters}, 45\penalty0 (13)\penalty0
  6400--6407, {\bf 2018{\natexlab{a}}}.

\bibitem[Filacchione et~al.(2018{\natexlab{b}})Filacchione, Ciarniello,
  D'Aversa, Capaccioni, Cerroni, Buratti, Clark, Stephan, and
  Plainaki]{Filacchione2018}
Filacchione G. and~8~colleagues.
\newblock \href{http://doi.org/10.1002/2017GL076869}{Photometric {{Modeling}}
  and {{VIS}}-{{IR Albedo Maps}} of {{Dione From Cassini}}-{{VIMS}}}.
\newblock \emph{Geophysical Research Letters}, 45\penalty0 (5)\penalty0
  2184--2192, {\bf 2018{\natexlab{b}}}.

\bibitem[Filacchione et~al.(2016)Filacchione, D'Aversa, Capaccioni, Clark,
  Cruikshank, Ciarniello, Cerroni, Bellucci, Brown, Buratti, Nicholson,
  Jaumann, McCord, Sotin, Stephan, and Dalle~Ore]{Filacchione2016}
Filacchione G. and~15~colleagues.
\newblock \href{http://doi.org/10.1016/j.icarus.2016.02.019}{Saturn's icy
  satellites investigated by {{Cassini}}-{{VIMS}}. {{IV}}. {{Daytime}}
  temperature maps}.
\newblock \emph{Icarus}, 271\penalty0 292--313, {\bf 2016}.

\bibitem[Fink and Larson(1975)]{Fink1975}
Fink U. and Larson H.~P.
\newblock \href{http://doi.org/10.1016/0019-1035(75)90058-5}{Temperature
  dependence of the water-ice spectrum between 1 and 4 microns: {{Application}}
  to {{Europa}}, {{Ganymede}} and {{Saturn}}'s rings}.
\newblock \emph{Icarus}, 24\penalty0 (4)\penalty0 411--420, {\bf 1975}.

\bibitem[Grundy and Schmitt(1998)]{Grundy1998}
Grundy W.~M. and Schmitt B.
\newblock \href{http://doi.org/10.1029/98JE00738}{The temperature-dependent
  near-infrared absorption spectrum of hexagonal {{H2O}} ice}.
\newblock \emph{Journal of Geophysical Research: Planets}, 103\penalty0
  (E11)\penalty0 25809--25822, {\bf 1998}.

\bibitem[Grundy et~al.(1999)Grundy, Buie, Stansberry, Spencer, and
  Schmitt]{Grundy1999}
Grundy W.~M. and~4~colleagues.
\newblock \href{http://doi.org/10.1006/icar.1999.6216}{Near-{{Infrared
  Spectra}} of {{Icy Outer Solar System Surfaces}}: {{Remote Determination}} of
  {{H2O Ice Temperatures}}}.
\newblock \emph{Icarus}, 142\penalty0 (2)\penalty0 536--549, {\bf 1999}.

\bibitem[Hansen et~al.(2006)Hansen, Esposito, Stewart, Colwell, Hendrix, Pryor,
  Shemansky, and West]{Hansen2006}
Hansen C.~J. and~7~colleagues.
\newblock \href{http://doi.org/10.1126/science.1121254}{Enceladus' {{Water
  Vapor Plume}}}.
\newblock \emph{Science}, 311\penalty0 (5766)\penalty0 1422--1425, {\bf 2006}.

\bibitem[Hansen and McCord(2004)]{Hansen2004}
Hansen G.~B. and McCord T.~B.
\newblock \href{http://doi.org/10.1029/2003JE002149}{Amorphous and crystalline
  ice on the {{Galilean}} satellites: {{A}} balance between thermal and
  radiolytic processes}.
\newblock \emph{Journal of Geophysical Research: Planets}, 109\penalty0 (E1),
  {\bf 2004}.

\bibitem[Hapke(1981)]{Hapke1981}
Hapke B.
\newblock \href{http://doi.org/10.1029/JB086iB04p03039}{Bidirectional
  reflectance spectroscopy: 1. {{Theory}}}.
\newblock \emph{Journal of Geophysical Research: Solid Earth}, 86\penalty0
  (B4)\penalty0 3039--3054, {\bf 1981}.

\bibitem[Hapke(2012)]{Hapke2012}
Hapke B.
\newblock \href{http://doi.org/10.1017/CBO9781139025683}{\emph{Theory of
  {{Reflectance}} and {{Emittance Spectroscopy}}}}.
\newblock {Cambridge University Press}, second edition, {\bf 2012}.

\bibitem[Hapke(1963)]{Hapke1963}
Hapke B.~W.
\newblock \href{http://doi.org/10.1029/JZ068i015p04571}{A theoretical
  photometric function for the lunar surface}.
\newblock \emph{Journal of Geophysical Research}, 68\penalty0 (15)\penalty0
  4571--4586, {\bf 1963}.

\bibitem[Jaumann et~al.(2008)Jaumann, Stephan, Hansen, Clark, Buratti, Brown,
  Baines, Newman, Bellucci, Filacchione, Coradini, Cruikshank, Griffith,
  Hibbitts, McCord, Nelson, Nicholson, Sotin, and Wagner]{Jaumann2008}
Jaumann R. and~18~colleagues.
\newblock \href{http://doi.org/10.1016/j.icarus.2007.09.013}{Distribution of
  icy particles across {{Enceladus}}' surface as derived from
  {{Cassini}}-{{VIMS}} measurements}.
\newblock \emph{Icarus}, 193\penalty0 (2)\penalty0 407--419, {\bf 2008}.

\bibitem[Kreslavsky et~al.(2000)Kreslavsky, Shkuratov, Velikodsky, Kaydash,
  Stankevich, and Pieters]{Kreslavsky2000}
Kreslavsky M.~A. and~5~colleagues.
\newblock \href{http://doi.org/10.1029/1999JE001150}{Photometric properties of
  the lunar surface derived from {{Clementine}} observations}.
\newblock \emph{Journal of Geophysical Research: Planets}, 105\penalty0
  (E8)\penalty0 20281--20295, {\bf 2000}.

\bibitem[Le~Mou{\'e}lic et~al.(2019)Le~Mou{\'e}lic, Cornet, Rodriguez, Sotin,
  Seignovert, Barnes, Brown, Baines, Buratti, Clark, Nicholson, Lasue, Pasek,
  and Soderblom]{LeMouelic2019}
Le~Mou{\'e}lic S. and~13~colleagues.
\newblock \href{http://doi.org/10.1016/j.icarus.2018.09.017}{The {{Cassini
  VIMS}} archive of {{Titan}}: {{From}} browse products to global infrared
  color maps}.
\newblock \emph{Icarus}, 319\penalty0 121--132, {\bf 2019}.

\bibitem[Minnaert(1941)]{Minnaert1941}
Minnaert M.
\newblock \href{http://doi.org/10.1086/144279}{The reciprocity principle in
  lunar photometry}.
\newblock \emph{The Astrophysical Journal}, 93\penalty0 403--410, {\bf 1941}.

\bibitem[Newman et~al.(2008)Newman, Buratti, Brown, Jaumann, Bauer, and
  Momary]{Newman2008}
Newman S.~F. and~5~colleagues.
\newblock \href{http://doi.org/10.1016/j.icarus.2007.04.019}{Photometric and
  spectral analysis of the distribution of crystalline and amorphous ices on
  {{Enceladus}} as seen by {{Cassini}}}.
\newblock \emph{Icarus}, 193\penalty0 (2)\penalty0 397--406, {\bf 2008}.

\bibitem[Nicholson et~al.(2019)Nicholson, Ansty, Hedman, Creel, Ahlers,
  Harbison, Brown, Clark, Baines, Buratti, Sotin, and Badman]{Nicholson2019}
Nicholson P.~D. and~11~colleagues.
\newblock \href{http://doi.org/10.1016/j.icarus.2019.06.017}{Occultation
  observations of {{Saturn}}'s rings with {{Cassini VIMS}}}.
\newblock \emph{Icarus}, {\bf 2019}.

\bibitem[Oancea et~al.(2012)Oancea, Grasset, Le~Menn, Bollengier, Bezacier,
  Le~Mou{\'e}lic, and Tobie]{Oancea2012}
Oancea A. and~6~colleagues.
\newblock \href{http://doi.org/10.1016/j.icarus.2012.09.020}{Laboratory
  infrared reflection spectrum of carbon dioxide clathrate hydrates for
  astrophysical remote sensing applications}.
\newblock \emph{Icarus}, 221\penalty0 (2)\penalty0 900--910, {\bf 2012}.

\bibitem[Porco et~al.(2006)Porco, Helfenstein, Thomas, Ingersoll, Wisdom, West,
  Neukum, Denk, Wagner, Roatsch, Kieffer, Turtle, McEwen, Johnson, Rathbun,
  Veverka, Wilson, Perry, Spitale, Brahic, Burns, DelGenio, Dones, Murray, and
  Squyres]{Porco2006}
Porco C.~C. and~24~colleagues.
\newblock \href{http://doi.org/10.1126/science.1123013}{Cassini {{Observes}}
  the {{Active South Pole}} of {{Enceladus}}}.
\newblock \emph{Science}, 311\penalty0 (5766)\penalty0 1393--1401, {\bf 2006}.

\bibitem[Ries and Janssen(2015)]{Ries2015}
Ries P.~A. and Janssen M.
\newblock \href{http://doi.org/10.1016/j.icarus.2015.04.030}{A large-scale
  anomaly in {{Enceladus}}' microwave emission}.
\newblock \emph{Icarus}, 257\penalty0 88--102, {\bf 2015}.

\bibitem[Schenk et~al.(2011)Schenk, Hamilton, Johnson, McKinnon, Paranicas,
  Schmidt, and Showalter]{Schenk2011}
Schenk P. and~6~colleagues.
\newblock \href{http://doi.org/10.1016/j.icarus.2010.08.016}{Plasma, plumes and
  rings: {{Saturn}} system dynamics as recorded in global color patterns on its
  midsize icy satellites}.
\newblock \emph{Icarus}, 211\penalty0 (1)\penalty0 740--757, {\bf 2011}.

\bibitem[Schmitt et~al.(1998)Schmitt, Quirico, Trotta, and Grundy]{Schmitt1998}
Schmitt B. and~3~colleagues.
\newblock \href{http://doi.org/10.1007/978-94-011-5252-5_9}{Optical
  {{Properties}} of {{Ices From UV}} to {{Infrared}}}.
\newblock In \emph{Solar {{System Ices}}}, pp 199--240. {Springer Netherlands},
  {\bf 1998}.

\bibitem[Schr{\"o}der et~al.(2013)Schr{\"o}der, Mottola, Keller, Raymond, and
  Russell]{Schroder2013}
Schr{\"o}der S.~E. and~4~colleagues.
\newblock \href{http://doi.org/10.1016/j.pss.2013.06.009}{Resolved photometry
  of {{Vesta}} reveals physical properties of crater regolith}.
\newblock \emph{Planetary and Space Science}, 85\penalty0 198--213, {\bf 2013}.

\bibitem[Schr{\"o}der et~al.(2017)Schr{\"o}der, Mottola, Carsenty, Ciarniello,
  Jaumann, Li, Longobardo, Palmer, Pieters, Preusker, Raymond, and
  Russell]{Schroder2017}
Schr{\"o}der S.~E. and~11~colleagues.
\newblock \href{http://doi.org/10.1016/j.icarus.2017.01.026}{Resolved
  spectrophotometric properties of the {{Ceres}} surface from {{Dawn Framing
  Camera}} images}.
\newblock \emph{Icarus}, 288\penalty0 201--225, {\bf 2017}.

\bibitem[Scipioni et~al.(2017)Scipioni, Schenk, Tosi, D'Aversa, Clark, Combe,
  and Ore]{Scipioni2017}
Scipioni F. and~6~colleagues.
\newblock \href{http://doi.org/10.1016/j.icarus.2017.02.012}{Deciphering
  sub-micron ice particles on {{Enceladus}} surface}.
\newblock \emph{Icarus}, 290\penalty0 183--200, {\bf 2017}.

\bibitem[Shkuratov et~al.(2011)Shkuratov, Kaydash, Korokhin, Velikodsky,
  Opanasenko, and Videen]{Shkuratov2011}
Shkuratov Y. and~5~colleagues.
\newblock \href{http://doi.org/10.1016/j.pss.2011.06.011}{Optical measurements
  of the {{Moon}} as a tool to study its surface}.
\newblock \emph{Planetary and Space Science}, 59\penalty0 (13)\penalty0
  1326--1371, {\bf 2011}.

\bibitem[Shkuratov et~al.(1999)Shkuratov, Kreslavsky, Ovcharenko, Stankevich,
  Zubko, Pieters, and Arnold]{Shkuratov1999}
Shkuratov Y.~G. and~6~colleagues.
\newblock \href{http://doi.org/10.1006/icar.1999.6154}{Opposition {{Effect}}
  from {{Clementine Data}} and {{Mechanisms}} of {{Backscatter}}}.
\newblock \emph{Icarus}, 141\penalty0 (1)\penalty0 132--155, {\bf 1999}.

\bibitem[Smith et~al.(1982)Smith, Soderblom, Batson, Bridges, Inge, Masursky,
  Shoemaker, Beebe, Boyce, Briggs, Bunker, Collins, Hansen, Johnson, Mitchell,
  Terrile, Cook, Cuzzi, Pollack, Danielson, Ingersoll, Davies, Hunt, Morrison,
  Owen, Sagan, Veverka, Strom, and Suomi]{Smith1982}
Smith B.~A. and~28~colleagues.
\newblock \href{http://doi.org/10.1126/science.215.4532.504}{A {{New Look}} at
  the {{Saturn System}}: {{The Voyager}} 2 {{Images}}}.
\newblock \emph{Science}, 215\penalty0 (4532)\penalty0 504--537, {\bf 1982}.

\bibitem[Spahn et~al.(2006)Spahn, Schmidt, Albers, H{\"o}rning, Makuch, Sei\ss,
  Kempf, Srama, Dikarev, Helfert, {Moragas-Klostermeyer}, Krivov, Srem{\v
  c}evi{\'c}, Tuzzolino, Economou, and Gr{\"u}n]{Spahn2006}
Spahn F. and~15~colleagues.
\newblock \href{http://doi.org/10.1126/science.1121375}{Cassini {{Dust
  Measurements}} at {{Enceladus}} and {{Implications}} for the {{Origin}} of
  the {{E Ring}}}.
\newblock \emph{Science}, 311\penalty0 (5766)\penalty0 1416--1418, {\bf 2006}.

\bibitem[Spencer et~al.(2006)Spencer, Pearl, Segura, Flasar, Mamoutkine,
  Romani, Buratti, Hendrix, Spilker, and Lopes]{Spencer2006}
Spencer J.~R. and~9~colleagues.
\newblock \href{http://doi.org/10.1126/science.1121661}{Cassini {{Encounters
  Enceladus}}: {{Background}} and the {{Discovery}} of a {{South Polar Hot
  Spot}}}.
\newblock \emph{Science}, 311\penalty0 (5766)\penalty0 1401--1405, {\bf 2006}.

\bibitem[Squyres et~al.(1983)Squyres, Reynolds, Cassen, and Peale]{Squyres1983}
Squyres S.~W. and~3~colleagues.
\newblock \href{http://doi.org/10.1016/0019-1035(83)90152-5}{The evolution of
  {{Enceladus}}}.
\newblock \emph{Icarus}, 53\penalty0 (2)\penalty0 319--331, {\bf 1983}.

\bibitem[Taffin et~al.(2012)Taffin, Grasset, Le~Menn, Bollengier, Giraud, and
  Le~Mou{\'e}lic]{Taffin2012}
Taffin C. and~5~colleagues.
\newblock \href{http://doi.org/10.1016/j.pss.2011.08.015}{Temperature and grain
  size dependence of near-{{IR}} spectral signature of crystalline water ice:
  {{From}} lab experiments to {{Enceladus}}' south pole}.
\newblock \emph{Planetary and Space Science}, 61\penalty0 (1)\penalty0
  124--134, {\bf 2012}.

\bibitem[Verbiscer and Veverka(1994)]{Verbiscer1994}
Verbiscer A.~J. and Veverka J.
\newblock \href{http://doi.org/10.1006/icar.1994.1112}{A {{Photometric Study}}
  of {{Enceladus}}}.
\newblock \emph{Icarus}, 110\penalty0 (1)\penalty0 155--164, {\bf 1994}.

\bibitem[Waite et~al.(2006)Waite, Combi, Ip, Cravens, McNutt, Kasprzak, Yelle,
  Luhmann, Niemann, Gell, Magee, Fletcher, Lunine, and Tseng]{Waite2006}
Waite J.~H. and~13~colleagues.
\newblock \href{http://doi.org/10.1126/science.1121290}{Cassini {{Ion}} and
  {{Neutral Mass Spectrometer}}: {{Enceladus Plume Composition}} and
  {{Structure}}}.
\newblock \emph{Science}, 311\penalty0 (5766)\penalty0 1419--1422, {\bf 2006}.

\end{thebibliography}


\onecolumn
\appendix
\section{Supplementary material}
\setcounter{figure}{0}

\begin{landscape}
    \begin{figure*}[!ht]
        \vspace{1.75cm}
        \subcaptionbox{$\displaystyle
            I/F
            =
            \cos\left( \frac{\alpha}{2} \right)
            \cdot
            \cos\left[
                \frac{\pi}{\pi - \alpha} \left( \gamma - \frac{\alpha}{2} \right)
            \right]
            \cdot
            \frac{\left( \cos\beta \right)^{ \frac{\alpha}{\pi - \alpha} }}{\cos\gamma}
            \cdot
            \left( k_1 + k_2 \alpha \right)
        $}
        {\includegraphics[height=.25\linewidth]{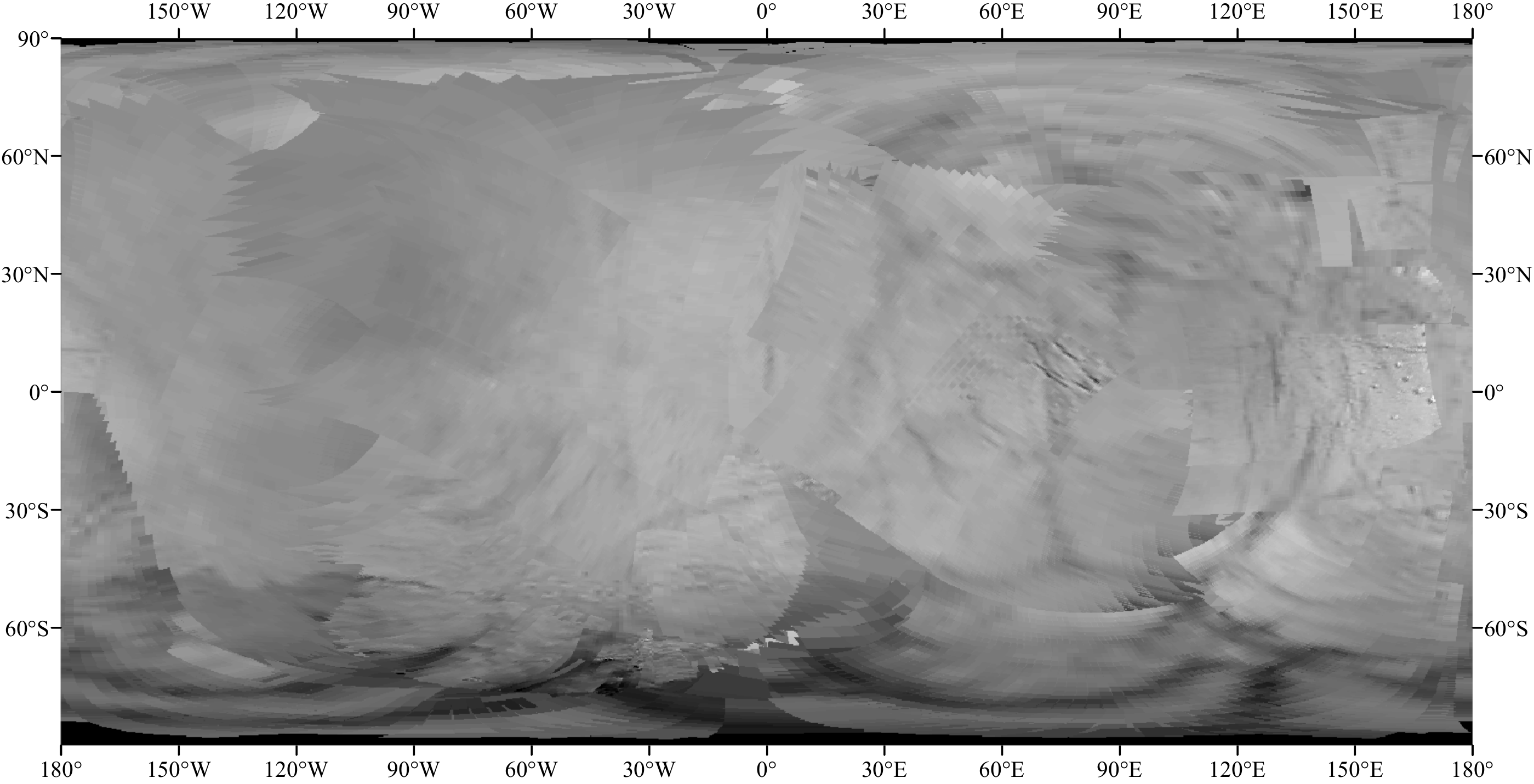}}
        \subcaptionbox{$\displaystyle
            I/F
            =
            \cos\left( \frac{\alpha}{2} \right)
            \cdot
            \cos\left[
                \frac{\pi}{\pi - \alpha} \left( \gamma - \frac{\alpha}{2} \right)
            \right]
            \cdot
            \frac{\left( \cos\beta \right)^{ \frac{\alpha}{\pi - \alpha} }}{\cos\gamma}
            \cdot
            k_1 \exp^{k_2 \alpha}
        $}
        {\includegraphics[height=.25\linewidth]{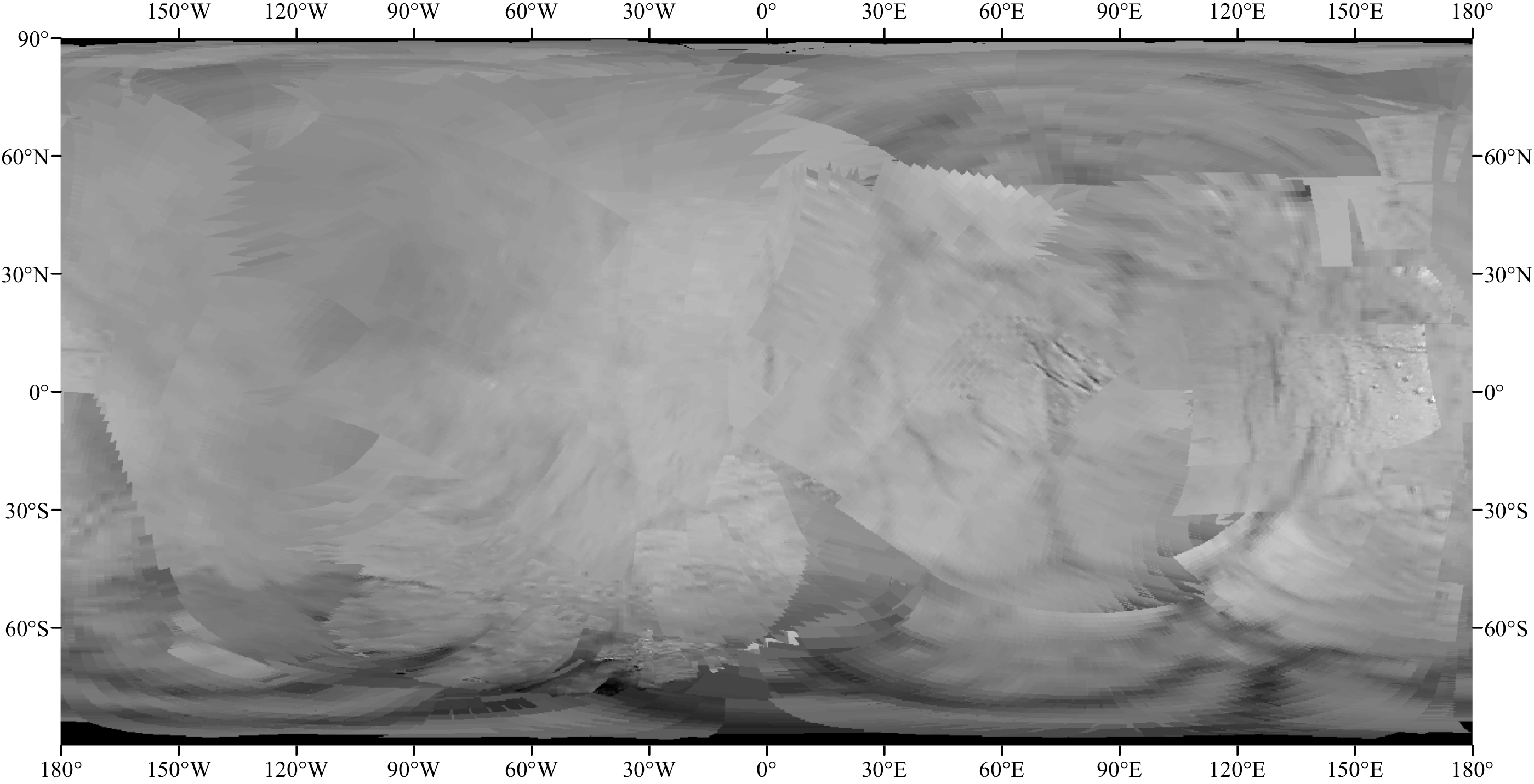}}\\
        \vspace{1cm}
        \subcaptionbox{$\displaystyle
            I/F
            =
            \cos\left( \frac{\alpha}{2} \right)
            \cdot
            \cos\left[
                \frac{\pi}{\pi - \alpha} \left( \gamma - \frac{\alpha}{2} \right)
            \right]
            \cdot
            \frac{\left( \cos\beta \right)^{ k \frac{\alpha}{\pi - \alpha} }}{\cos\gamma}
            \cdot
            \left( k_1 + k_2 \alpha \right)
        $}
        {\includegraphics[height=.25\linewidth]{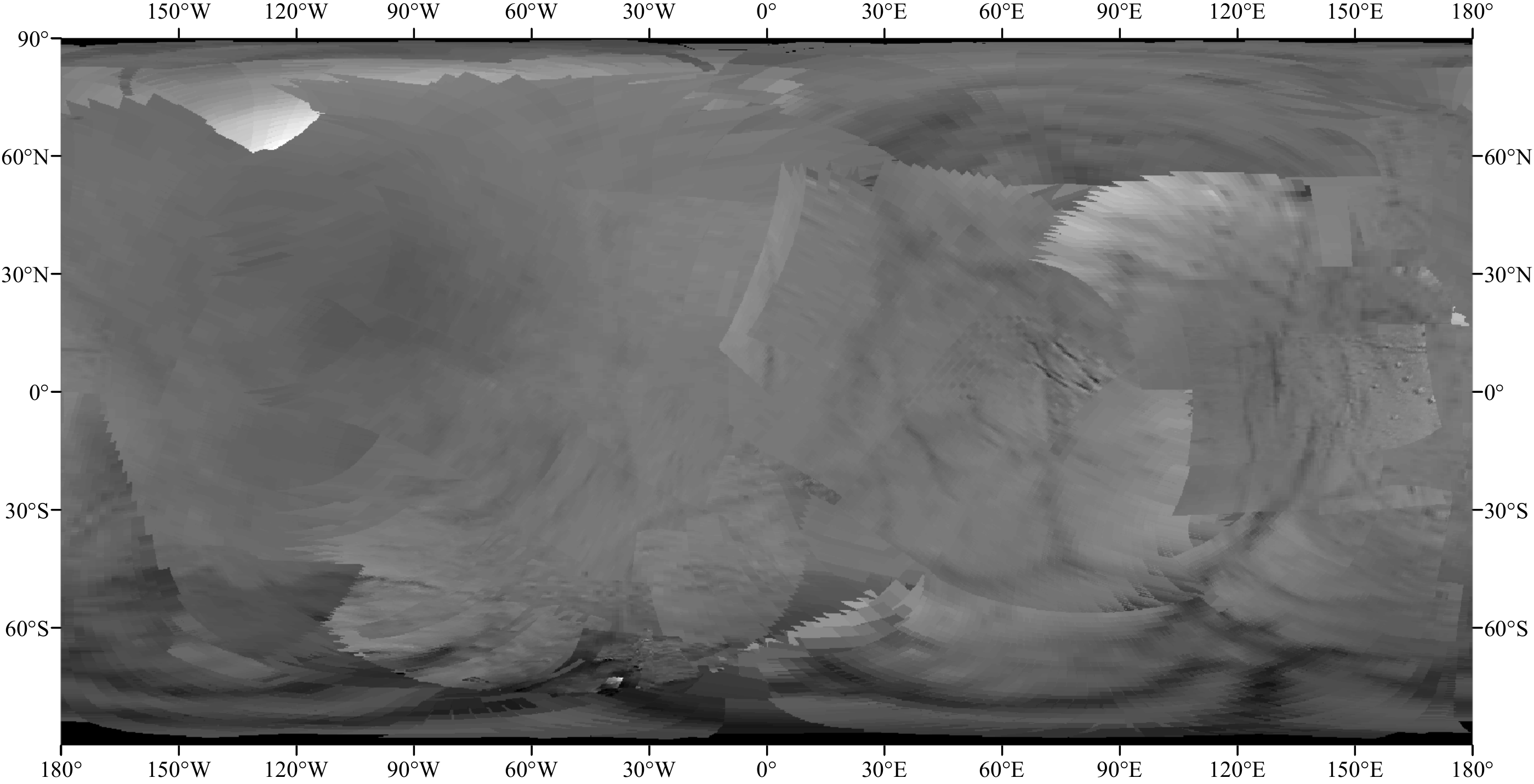}}
        \subcaptionbox{$\displaystyle
            I/F
            =
            \cos\left( \frac{\alpha}{2} \right)
            \cdot
            \cos\left[
                \frac{\pi}{\pi - \alpha} \left( \gamma - \frac{\alpha}{2} \right)
            \right]
            \cdot
            \frac{\left( \cos\beta \right)^{ k \frac{\alpha}{\pi - \alpha} }}{\cos\gamma}
            \cdot
            k_1 \exp^{k_2 \alpha}
        $}
        {\includegraphics[height=.25\linewidth]{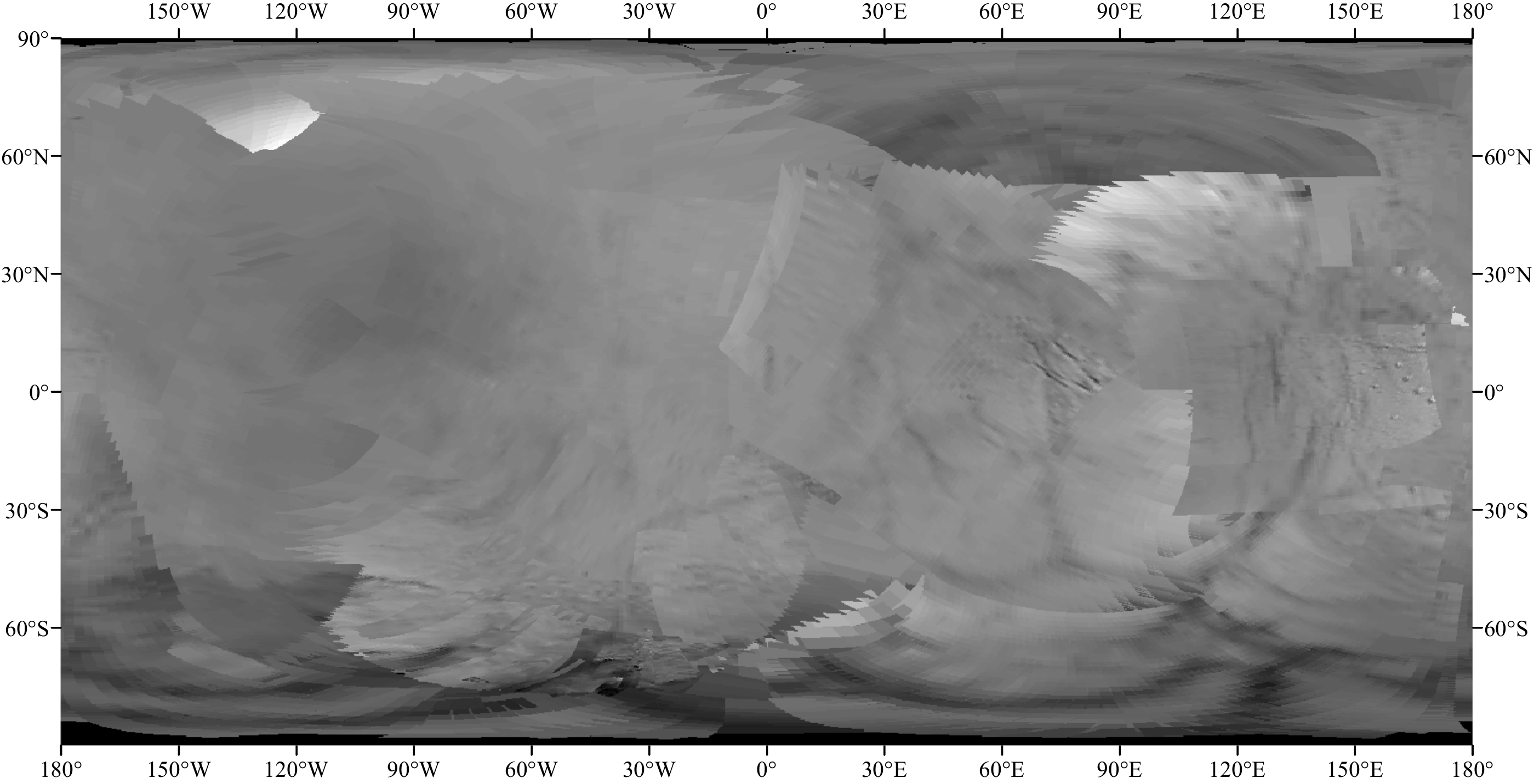}}\\
        \vspace{.5cm}
        \caption{Mosaic at \SI{1.8}{\um} corrected with (a) Akimov disk function and linear phase function, (b) Akimov disk function and exponential phase function, (c) parametrized Akimov disk function and linear phase function and (d) parametrized Akimov disk function and exponential phase function.}
        \label{fig:fig_S1}
    \end{figure*}
\end{landscape}

\begin{landscape}
    \begin{figure*}[!ht]
        \vspace{1.75cm}
        \subcaptionbox{$\displaystyle
            I/F
            =
            \mu_0^k \mu^{k-1}
            \cdot
            \left( k_1 + k_2 \alpha \right)
        $}
        {\includegraphics[height=.25\linewidth]{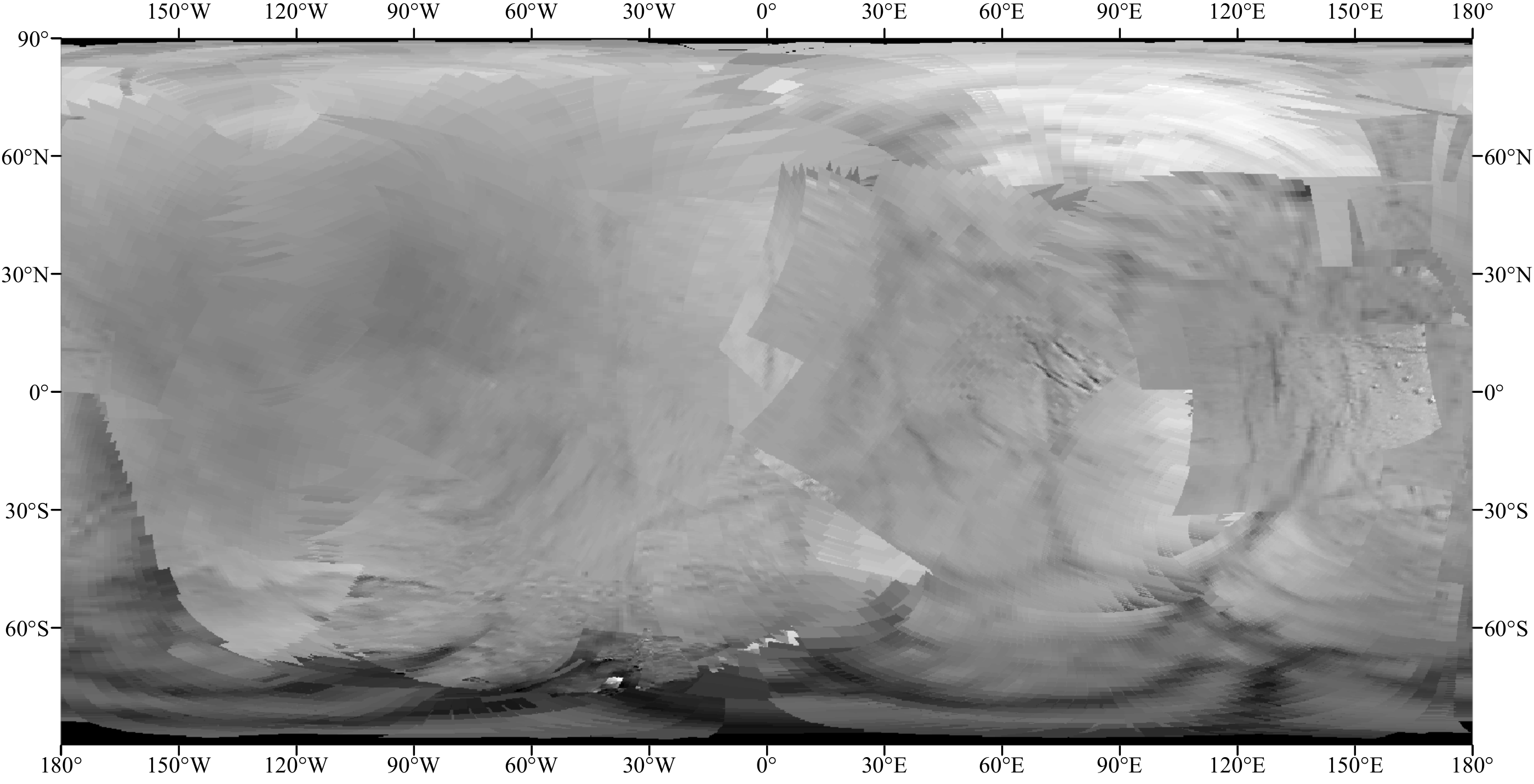}}
        \subcaptionbox{$\displaystyle
            I/F
            =
            \mu_0^k \mu^{k-1}
            \cdot
            k_1 \exp^{k_2 \alpha}
        $}
        {\includegraphics[height=.25\linewidth]{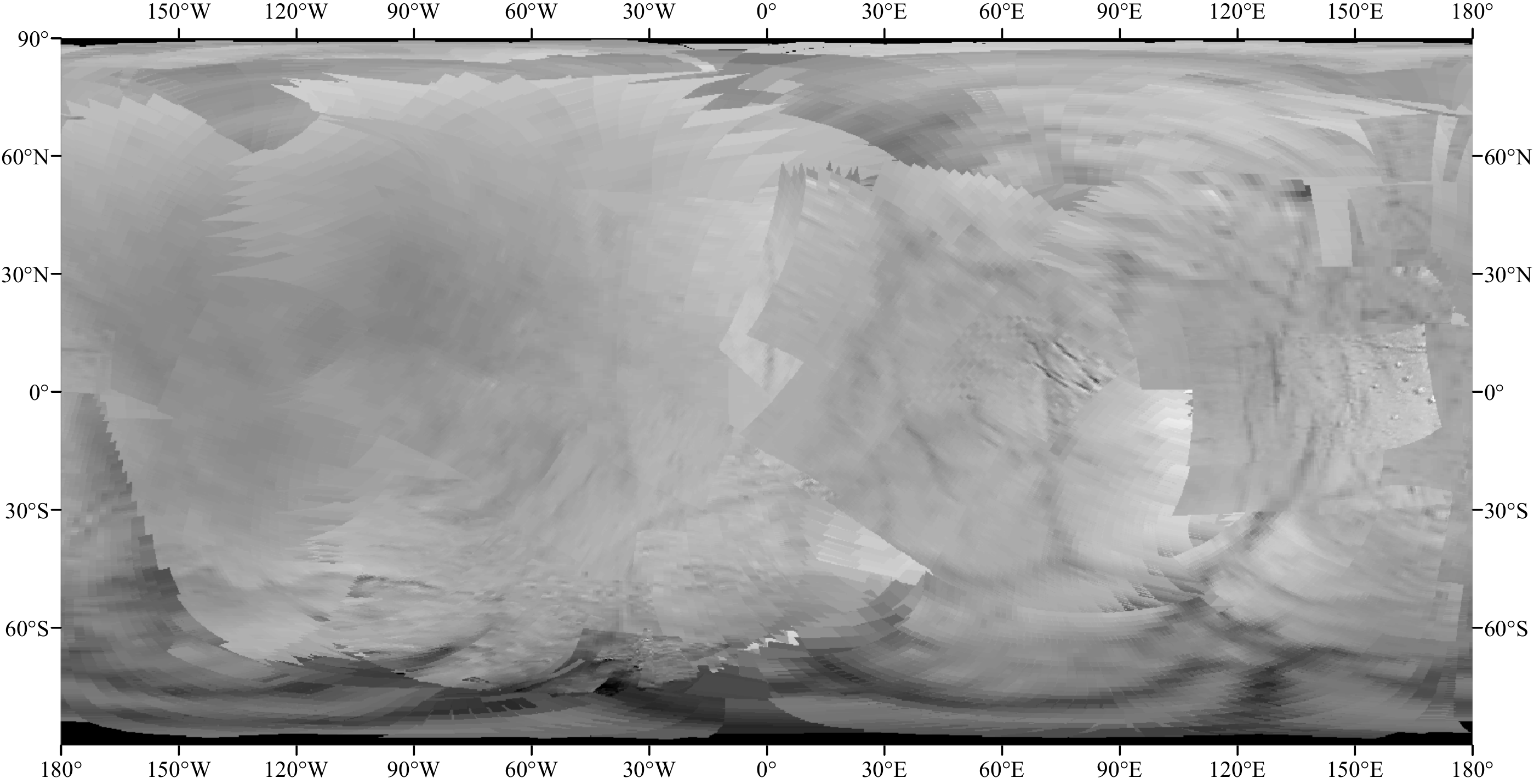}}\\
        \vspace{1cm}
        \subcaptionbox{$\displaystyle
            I/F
            =
            \left[
                k \frac{2 \mu_0}{\mu_0 + \mu} + \left( 1 - k \right) \mu_0
            \right]
            \cdot
            \left( k_1 + k_2 \alpha \right)
        $}
        {\includegraphics[height=.25\linewidth]{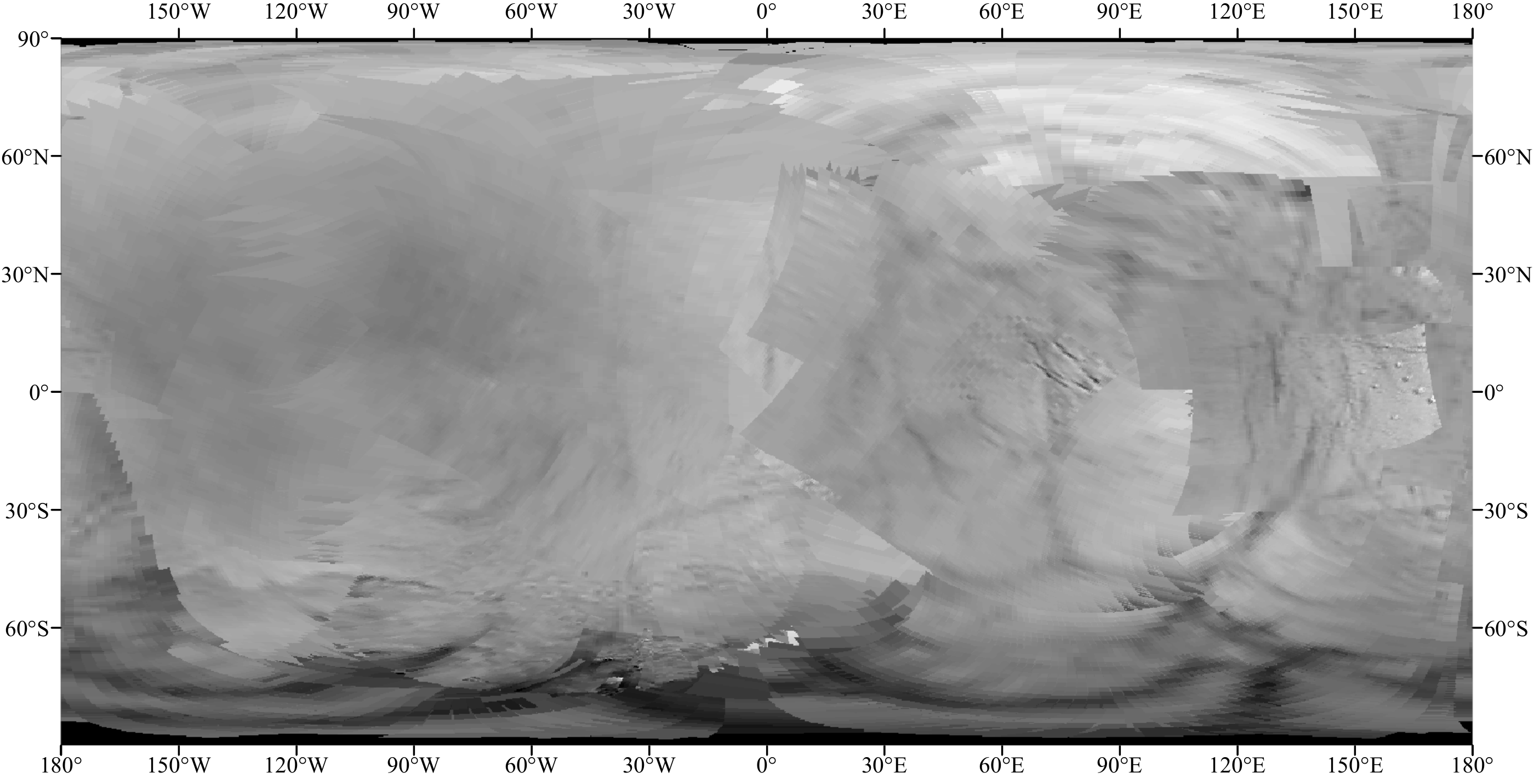}}
        \subcaptionbox{$\displaystyle
            I/F
            =
            \left[
                k \frac{2 \mu_0}{\mu_0 + \mu} + \left( 1 - k \right) \mu_0
            \right]
            \cdot
            k_1 \exp^{k_2 \alpha}
        $}
        {\includegraphics[height=.25\linewidth]{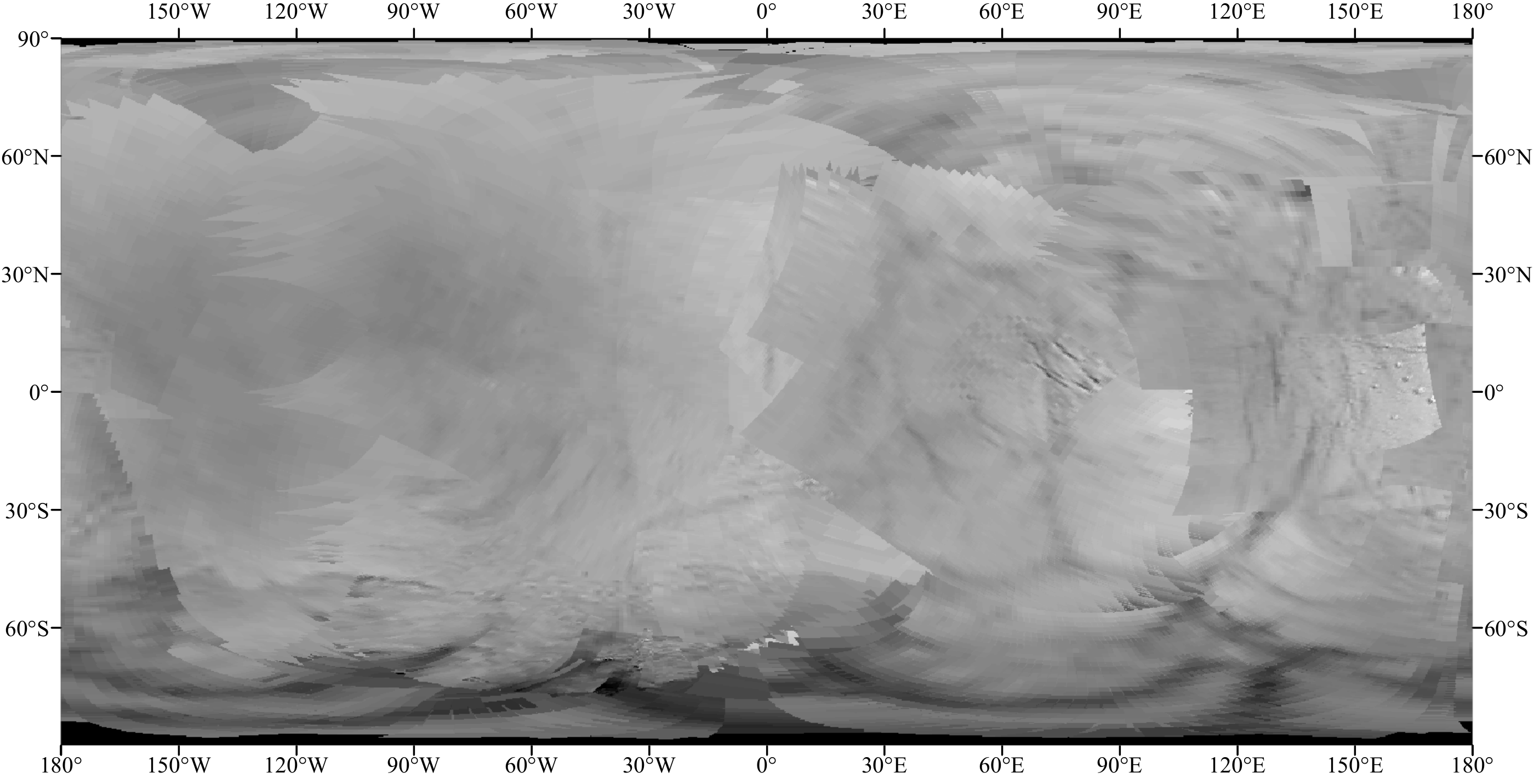}}\\
        \vspace{.5cm}
        \caption{Mosaic at \SI{1.8}{\um} corrected with (a) Minnaert disk function and linear phase function, (b) Minnaert disk function and exponential phase function, (c) L-S/Lambert disk function and linear phase function and (d) L-S/Lambert disk function and exponential phase.}
        \label{fig:fig_S2}
    \end{figure*}
\end{landscape}

This supplementary material aims to provide the reader with additional information on the different photometric functions that have been tested during the study. We used non-linear least squares to fit the photometric functions to data. The tested functions are summarized in the following table (\tabref{tab_S1}). The standard deviation errors correspond to the square root of the estimated covariance matrix diagonal. The corresponding parameters were computed by minimizing the standard deviation error. The resulting mosaics are presented in \figref{fig_S1} and \figref{fig_S2} for one specific wavelength (\SI{1.8}{\um}). The first figure (\figref{fig_S1}) illustrates the difference between two models of disk functions (Akimov and parametrized Akimov). It also highlights the differences regarding the phase function, whether we choose a linear phase function or an exponential one. The complete photometric function is mentioned above the corresponding map. The second figure (\figref{fig_S2}) shows two other disk functions (Lommel-Seeliger/Lambert and Minnaert models) combined with linear and exponential phase functions.
Please note that we took the disk function parameter $k$ constant with phase. A possibility for future studies would be to consider $k$ linearly dependent with the phase as proposed by \cite{Schroder2017}.

We notice that the association of the parametrized Akimov disk function with a linear phase function provides the best standard deviation errors for the computed parameters. However, the fit improvement remains small. Moreover, the resulting mosaic is downgraded relative to the one corrected for the association of the Akimov disk function with the linear phase function. The exponential phase function does improve the correction on some parts but deteriorates some others.

\vspace{.5cm}

\begin{table*}[!ht]
    \caption{Coefficients for the disk function and phase function parameters, computed for $I/F$ at \SI{1.8}{\um}. Parameters are valid for $(i, e) < \ang{80}$ and $\alpha < \ang{130}$. We evaluate the performance of photometric models by calculating the standard deviation errors.}
    \label{tab:tab_S1}
    \begin{tabular}{l l r r r r r r}
        \toprule
        &  & \multicolumn{3}{c}{Parameters} & \multicolumn{3}{c}{Standard deviation errors} \\
        \cmidrule{3-8}
        Disk function       & Phase function & \multicolumn{1}{c}{$k$} & \multicolumn{1}{c}{$k_1$} & \multicolumn{1}{c}{$k_2$}  &  \multicolumn{1}{c}{$\sigma(k)$} & \multicolumn{1}{c}{$\sigma(k_1)$} & \multicolumn{1}{c}{$\sigma(k_2)$} \\
        \midrule
        \multirow{2}{*}{Akimov}              & Linear         &       & 0.698 & -0.250 &              & \num{5.7e-4} & \num{8.5e-4} \\
                                             & Exponential    &       & 0.716 & -0.464 &              & \num{7.5e-4} & \num{1.8e-3} \\
        \cmidrule{3-8}
        \multirow{2}{*}{Parametrized Akimov} & Linear         & 2.422 & 0.717 & -0.243 & \num{1.3e-2} & \num{5.5e-4} & \num{8.0e-4} \\
                                             & Exponential    & 2.421 & 0.731 & -0.427 & \num{1.3e-2} & \num{6.9e-4} & \num{1.7e-3} \\
        \cmidrule{3-8}
        \multirow{2}{*}{Minnaert}            & Linear         & 0.741 & 0.806 & -0.340 & \num{7.7e-4} & \num{6.8e-4} & \num{7.9e-4} \\
                                             & Exponential    & 0.748 & 0.860 & -0.619 & \num{7.6e-4} & \num{8.8e-4} & \num{1.5e-3} \\
        \cmidrule{3-8}
        \multirow{2}{*}{L-S/Lambert}         & Linear         & 0.421 & 0.813 & -0.333 & \num{1.6e-3} & \num{7.0e-4} & \num{8.2e-4} \\
                                             & Exponential    & 0.406 & 0.863 & -0.592 & \num{1.5e-3} & \num{9.0e-4} & \num{1.5e-3} \\
        \bottomrule
    \end{tabular}
\end{table*}

\end{document}